\documentclass[prb,onecolumn,preprintnumbers,amsmath,amssymb]{revtex4}
\usepackage{graphicx}
\begin{document}

\title{ Exciton correlations and input-output relations in non-equilibrium  exciton  superfluids }
\author{  Jinwu Ye $^{1,2}$, Fadi Sun$^{1,3}$, Yi-Xiang Yu $^{1,3}$,  and Wuming Liu $^{3}$}
\affiliation{
$^{1}$ Department of Physics and Astronomy,
Mississippi State University, MS, 39762, USA  \\
$^{2}$ Department of Physics, Capital Normal University, Beijing,
100048 China  \\
$^{3}$Institute of Physics, Chinese Academy of Sciences,
Beijing, 100080, China }
\date{\today }
Corresponding author Jinwu Ye's: E-mail: jy306@ccs.msstate.edu~~
Phone: 662-325-2926 ~~
Fax No. 662-325-8898
\begin{abstract}
  The Photoluminescent (PL)  measurements on photons and the transport measurements on excitons are the two types of independent and complementary
 detection tools to search for  possible exciton superfluids  in electron-hole semi-conductor bilayer systems. In fact,
 it was believed  that the transport measurements can provide more direct evidences on superfluids than the spectroscopic measurements.
 It is important to establish the relations between the two kinds of measurements.
 In this paper, using quantum Heisenberg-Langevin equations, we establish such a connection by calculating various exciton correlation functions
 in the putative exciton superfluids. These correlation functions include both normal and anomalous
 Greater, Lesser, Advanced, Retarded, and Time-ordered  exciton Green functions and also various two exciton correlation functions.
 We also evaluate the corresponding normal and anomalous spectral weights and the Keldysh distribution functions.
 We stress the violations of the fluctuation and dissipation theorem among these various exciton correlation functions
 in the non-equilibrium exciton superfluids.
 We also explore the input-output relations between various exciton correlation
 functions and those of emitted photons such as  the angle resolved photon power spectrum, phase sensitive two mode squeezing spectrum and two photon correlations.
 Applications to possible superfluids in the exciton-polariton systems are also mentioned.
 For a comparison, using conventional imaginary time formalism, we also calculate all the exciton correlation functions in an
 equilibrium dissipative exciton superfluid in the  electron-electron coupled semi-conductor bilayers at
 the quantum Hall regime at the total filling factor $ \nu_T= 1 $.
 We stress the analogies and also important differences between the correlations functions in the two exciton superfluid systems.
\end{abstract}

\maketitle

\section{ Introduction }

An exciton is a bound state of an electron and a hole in the band structure of a solid.
When excitons are sufficiently apart from each other, they may behave as bosons.
Although Bose-Einstein condensation (BEC) of excitons was proposed
more than 3 decades ago \cite{blatt,kohn}, no exciton superfluid phase has been
observed in any bulk solids yet. Recently, degenerate exciton
systems  in quasi-two-dimensional semiconductor GaAs/AlGaAs
electron-hole coupled bilayers (EHBL) \cite{loz} have been produced by many experimental groups with
photo-generated  method \cite{butov} or gate-voltage generated method \cite{field,camb} and electron-electron coupled
bilayers at quantum Hall regime at the total filling factor $ \nu_T
= 1 $ (BLQH) \cite{blqhexp,counterflow,psdw}.
It was widely believed that EHBL and BLQH are two of
the most promising systems to observe BEC of excitons among any
solid state systems.  Indeed, there have been extensive experimental searches for exciton superfluids in both systems \cite{butov,field,expol,blqhexp}.
Now it was more or less established that an exciton superfluid subject to substantial dissipations has been observed in BLQH system \cite{blqhexp,counterflow,psdw}.
But there are still no convincing experimental evidences that the exciton superfluids (ESF) have been formed in EHBL at the present experimental conditions \cite{butov,field,camb}. In the photo-generated electron-hole bilayer (EHBL) samples \cite{butov}, various kinds of photoluminescence (PL) measurements
were made by different groups \cite{butov}. More recently, taking advantage of the long lifetimes of the in-direct excitons, researchers started to be able to  manipulate the excitons movements
and perform various transport properties of the excitons  \cite{tran1,tran2,tran3}. In the undoped electron-hole bilayer (EHBL) samples \cite{field,camb} which is a heterostructures insulated-gate field effect transistors, separate gates can be connected to electron layer and hole layer, so the densities of electron and holes can be tuned independently by varying the gate voltages.
Transport properties such as the Coulomb drag or counterflow can be performed in this experimental set-up. The PL signals, although weaker than those
in the  photo-generated samples, is still measurable. The transport measurements in \cite{field,camb,tran1,tran2,tran3} and the PL measurements in \cite{butov}
are complementary to each other.  It is important to search for exciton superfluid from both experimental methods.
Any signatures of the exciton condensations should show up in both type of experiments. In fact, it was believed that
the transport measurements can provide more direct evidences on superfluids than the spectroscopic measurements.
Indeed, it is all the peculiar transport properties which proved the existence of superfluid in the liquid $^{4} He $ at very low temperatures.

 There are previous theoretical works studying various photon emission spectra from  the putative ESF \cite{power,squ,excitonye}.
 In \cite{power,squ}, assuming the ESF has been formed in the EHBL, the authors studied the angle resolved photon
 spectrum, momentum distribution curve, energy distribution curve, two mode phase sensitive squeezing spectrum and two photon correlation functions
 from such an ESF. They found these photoluminescence (PL) display many unique and unusual features not shared by any other atomic or
 condensed matter systems. Observing all these salient features by  possible future angle resolved power spectrum,  phase sensitive homodyne experiment and HanburyBrown-Twiss type of experiment could lead to conclusive evidences of exciton superfluid in these systems.
 However, all the previous theoretical works focused on the the PL of the emitted photons,
 but various properties of the excitons themselves have not been addressed.
 There are also some previous theoretical work\cite{bcsdrag,bcsdrag2} on the Coulomb drag in electron-hole
 bilayer in the BCS side by using weak coupling mean field BCS theory. However, as analyzed in \cite{power,squ}, for the
 experimental relevant density regime $ n \sim 10^{10} cm^{-2} $, the excitons are tightly bound pairs in real space, so all the experiments are
 in the BEC regime, so it remains interesting to study the exciton correlations in the strong  coupling BEC limit.

\begin{figure}
\includegraphics[width=6cm]{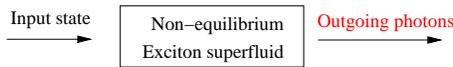}
\hspace{0.3cm}
\caption{ The irreversible photon emission process in an non-equilibrium exciton superfluid.
The Time reversal symmetry is broken.  The photons are emitted and never flow back, so the arrows only go to one direction.
The outgoing photons correlation functions are related to those of excitons by the input-output relations discussed in Sec.VI. As shown in Sec.IV and V, the fluctuation dissipation theorem (FDT) does not apply to this kind of quantum open systems.   }
\label{fig1}
\end{figure}

 As mentioned above, the  EHBL and BLQH are two of the most promising systems to observe BEC of excitons among any
 solid state systems. However, there are some crucial differences between the exciton superfluids in the two systems.
 The exciton + photon system in the EHBL is a typical quantum open system subject to a dissipative bath shown in the Fig.1.
 The photon emission process is a typical irreversible outgoing process breaking the time-reversal symmetry.
 However, the exciton superfluid in the BLQH at the total filling factor $ \nu_T = 1 $  is a typical equilibrium system, also subject
 to a dissipative bath shown in the Fig.2. There are reversible exchange process between the excitons and the dissipative gapless excitations.
 These processes preserve the time reversal symmetry.
 The relations and differences between the two kinds of systems have never been addressed before.
 As mentioned above also, the PL measurements and the transport measurements are two independent  and complementary measurements to search for exciton superfluids in the EHBL. The PL detects the properties of the emitted photons, while the transport measurements detect those of the excitons.
 The excitons and photons are coupled to each other through the interaction term in the Hamiltonian Eqn.\ref{first}.
 So it is important to study the relations between the photon correlation functions and those of excitons which will establish
 relations between the two kinds of measurements. These relations are also crucial in bridging the interconnections between
 the optical communications and the electronic signal processing required at the integrated circuit devices.
 Here, we will address these outstanding open problems.

 Specifically, using quantum Heisenberg-Langevin equations, we will study various exciton correlation functions
 in the steady photon emission state. We will also investigate various input-output relations between these exciton correlation functions
 and those of emitting photons. The results achieved on exciton correlation functions are directly relevant to
 the transport properties of the excitons. They are also needed to evaluate the superfluid density and the critical
 velocity of a non-equilibrium exciton superfluid in Fig.1 and equilibrium exciton superfluid in Fig.2.
 The input-output relations can also be used to establish the connections between
 the transport measurements in \cite{tran1,tran2,tran3,field,camb} and the PL measurements in \cite{butov}.
 We will also explore the analogies and differences between the exciton correlation functions in the non-equilibrium exciton superfluid systems shown
 in Fig.1 and those in the equilibrium dissipative exciton superfluid systems shown in Fig.2. Non-equilibrium physics remain poorly understood.
 So the results achieved in this manuscript in the context of non-equilibrium  exciton superfluid may lead to some general concepts
 applicable to various other non-equilibrium problems, they
 could also shed some lights on the equilibrium dissipative  exciton superfluid observed in BLQH systems \cite{blqhexp}.

\begin{figure}
\includegraphics[width=6cm]{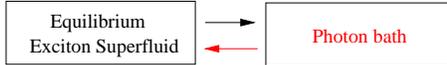}
\hspace{0.3cm}
\caption{ The reversible exchange processes between exciton superfluid and photon bath in an equilibrium exciton superfluid.
The Time reversal symmetry is not broken.  The photons are flowing back and forth, so the arrows go to both directions.
 As shown in Sec.IX, the fluctuation dissipation theorem (FDT) holds in
this kind of equilibrium systems. The dissipative exciton superfluids in BLQH \cite{blqhexp,counterflow,psdw} is in this class with the photon bath replaced by
gapless electron-hole excitations ( see \cite{justify} ). }
\label{fig2}
\end{figure}


 In parallel to the experimental search for the exciton  superfluid in the EHBL, there are also extensive experimental
 activities to search for exciton-polariton superfluid inside a
 planar micro-cavity. Although exciton condensation in a single quantum well ( SQW )
 has not been observed so far, there are some evidences for the observation of Exciton-polariton condensation in SWQ
 enclosed  inside a planar microcavity \cite{yama1,expol,expolmode}.
  These evidences include macroscopic occupation of the ground
  state, spectral and spatial narrowing, a peak at zero momentum in
  the momentum distribution and spontaneous linear polarization of
  the light emission and so on.  One photon and two photon correlation functions  have been measured in the
  experiments respectively. We expect that the methods developed and results achieved in this paper should also apply
  to the exciton-polariton condensation inside a microcavity.

  The rest of the paper is organized as follows. In Sec.II, to be self-contained, we review the effective exciton-photon interaction Hamiltonian,
  the input state, the input-output relation  and the exciton decay rate already studied in \cite{power,squ}, but we will use a slightly different notations
  than  \cite{power,squ} which are consistent with the condensed matter notations in \cite{fetter}.
  Most importantly, we also write out explicitly the exciton operator which will be the starting point to calculate various exciton correlation functions
  in the following sections. In Sec.III, we calculate the one photon correlation functions, especially the one photon anomalous Green function
  which was not evaluated in \cite{power,squ}.
  The central results of this paper are presented in Sec. IV-IX. We compute various normal one photon correlation functions such as
  the greater $ G^{>}_n $, lesser $ G^{<}_n $, retarded $ G^{R}_n $, advanced $ G^{A}_n $, the time-ordered $ G^{T}_n $ and the Keldysh component
  $ G^{K}_n $ in Sec.IV,  various anomalous one photon correlation functions such as the greater $ G^{>}_a $, lesser $ G^{<}_a $,
  retarded $ G^{R}_a $, advanced $ G^{A}_a $, time-ordered $ G^{T}_a $ and the Keldysh component  $ G^{K}_a $in Sec.V.
  Then we explore various Input-output relations between the photon correlation functions and those of excitons.
  In Sec.VII, we explore the non-trivial relations  between the angle resolved photon power spectrum (ARPS) and
  the normal  spectral weight and the normal distribution function, between the
  two mode squeezing spectrum and the  anomalous spectral weight and anomalous  distribution function at the experimentally relevant values.
  In Sec.VIII, we calculate the two exciton correlation functions and study
  their connections and differences than those of photons which can be measured by HanburyBrown-Twiss type experiments.
  These two-exciton correlations functions are important to transport measurements on excitons.
  In Sec.IX, using the imaginary time formalism, we study all the exciton correlations functions of an equilibrium dissipative exciton
  superfluid subject to the photon dissipation bath in Fig.2. Then we compare with the results achieved in the previous sections
  on  non-equilibrium exciton superfluid in Fig.1.  We reach conclusions and possible future perspectives in the final section X.
  Throughout the paper, we did not consider the possible important effects of spins of electrons
  and holes which lead to the formation of the bright excitons with $ J=\pm 1 $ and the dark excitons \cite{spinthe}
  with $ J=\pm 3/2 $, the effects of the trap, finite thickness and  disorders in the sample. These facts will be briefly discussed in the conclusion.

\section{The Photon-exciton interaction, initial state and Input-out formalism }

The total Hamiltonian of the exciton + photon is the sum of exciton part with the dipole-dipole interactions,
the photon part and the coupling between the two parts $H_{t}=H_{ex}+H_{ph}+H_{int}$ where \cite{power,squ}:
\begin{eqnarray}
H_{ex} &=&\sum_{\vec{k}}(E_{\vec{k}}^{ex}-\mu )b_{\vec{k}}^{\dagger }b_{\vec{%
k}}+ \frac{1}{2 A} \sum_{\vec{k}\vec{p}\vec{q}}V_{d}(q)b_{\vec{k}-\vec{q}}^{\dagger }b_{%
\vec{p}+\vec{q}}^{\dagger }b_{\vec{p}}b_{\vec{k}}  \nonumber \\
H_{ph} &=&\sum_{k}\omega _{k} a_{k}^{\dagger }a_{k}  \nonumber \\
H_{int} & = & \sum_{k}[ig(k)a_{k}e^{i \mu t } b_{\vec{k}}^{\dagger }+h.c.]
\label{first}
\end{eqnarray}%
 where  $ A $ is the area of the EHBL, the exciton energy $ E_{\vec{k}}^{ex} = \vec{k}^{2}/2M +
 E_{g}-E_{b}$, the photon frequency
 $ \omega_{k}=v_{g}\sqrt{k_{z}^{2}+\vec{k}^{2}}$ where
 $v_{g}=c/\sqrt{\epsilon } $ with $ c $  the light speed in the
 vacuum and $\epsilon \sim 12$  the dielectric constant of $GaAs$,
 $ k=( \vec{k},k_z) $ is the 3 dimensional momentum,
 $V_{d}(\vec{q})$ is the dipole-dipole interaction between the
 excitons \cite{excitonye}, $V_{d}(\left\vert \vec{r}\right\vert \gg
d)=e^{2}d^{2}/\left\vert \vec{r}\right\vert ^{3}$ and $  V_{d}(q=0)
= \frac{ 2 \pi e^{2} d}{ \epsilon } $  where $ d $ is the interlayer distance leads to a capacitive term for
the density fluctuation \cite{psdw}. The $ g(k) \sim
\vec{\epsilon}_{k\lambda }\cdot \vec{D}_{k} \times L^{-1/2}_{z} $ is
the coupling between the exciton and the photons  where  $ \vec{\epsilon}_{k\lambda } $ is the photon polarization,
$ \vec{D}_{k} $ is the transition dipole moment and $ L_{z}
\rightarrow \infty $ is the normalization length along the $ z $
direction \cite{power}. In a frame rotating with the frequency $ \mu $, the Hamiltonian becomes
\begin{eqnarray}
H^{R}_{ex} &=&\sum_{\vec{k}}(E_{\vec{k}}^{ex}-\mu )b_{\vec{k}}^{\dagger }b_{\vec{%
k}}+ \frac{1}{2 A} \sum_{\vec{k}\vec{p}\vec{q}}V_{d}(q)b_{\vec{k}-\vec{q}}^{\dagger }b_{%
\vec{p}+\vec{q}}^{\dagger }b_{\vec{p}}b_{\vec{k}}  \nonumber \\
H^{R}_{ph} &=&\sum_{k} (\omega _{k} - \mu ) a_{k}^{\dagger }a_{k}  \nonumber \\
H^{R}_{int} & = & \sum_{k}[ig(k)a_{k} b_{\vec{k}}^{\dagger }+h.c.]
\label{firstr}
\end{eqnarray}%
  In the following except in Sec. IX, all the calculations will be done in this rotating frame.

In the dilute limit, the average distance between the excitons is large, so the $V_{d}$ is relatively weak, therefore we may apply
the standard Bogoliubov approximation.
In the Exciton superfluid (ESF) phase, one can decompose the exciton operator into the condensation
part and the quantum fluctuation part above the condensation:
\begin{equation}
b_{\vec{k}}=\sqrt{N}\delta _{\vec{k}0}+\tilde{b}_{\vec{k}}
\label{conden}
\end{equation}
   The linear term in $\tilde{b}_{k}$ is
eliminated by setting the chemical potential $ \mu =E_{0}^{ex}+
\bar{n} V_{d}(0)= ( E_g-E_b) + \bar{n} V_{d}(0) $.
In a stationary state, the $ \mu $ is kept fixed at this value.
Upto the quadratic terms, the exciton Hamiltonian $H_{ex}$ is simplified to:
\begin{equation}
H_{sf}=\sum_{\vec{k}}[(\epsilon _{\vec{k}}+V_{d}(\vec{k})\bar{n})\tilde{b}_{%
\vec{k}}^{\dagger }\tilde{b}_{\vec{k}}+(\frac{V_{d}(\vec{k})\bar{n}}{2}%
\tilde{b}_{\vec{k}}^{\dagger }\tilde{b}_{-\vec{k}}^{\dagger }+h.c.)]
\label{quad}
\end{equation}%
where the density of the condensate $\bar{n}=N/S$. It can be
diagonalized by Bogoliubov transformation:
$H_{sf}=E(0)+\sum_{\vec{k}}E(\vec{k})\beta _{\vec{k}}^{\dagger }\beta _{\vec{k%
}} $ where $
E(\vec{k})=\sqrt{\epsilon _{\vec{k}}[\epsilon _{\vec{k}}+2\bar{n}V_{d}(\vec{k%
})]} $ is the excitation spectrum, $ \beta _{\vec{k}}=u_{\vec{k}}\tilde{b}_{\vec{k}}+v_{\vec{k}}\tilde{b}_{-\vec{k%
}}^{\dagger } $ is the Bogoliubov quasi-particle annihilation
operators with the two coherence factors $ u_{\vec{k}}^{2} ( v_{\vec{k}}^{2} ) =\frac{\epsilon _{\vec{k}}+\bar{n}V_{d}(\vec{k})}{2E(\vec{k%
})} \pm \frac{1}{2} $.  As $ \vec{k}\rightarrow 0 $, $ E(\vec{k})=u \left\vert
\vec{k} \right\vert $ shown in Fig.3 where the velocity of the
quasi-particle is $u=\sqrt{ \bar{n}V_{d}(0)/M} $.
At $ T=0 $, the number of excitons out of the condensate is: $
n^{\prime}(T=0) = \frac{1}{S} \sum_{\vec{k}}
 \langle \tilde{b}^{\dagger}_{\vec{k}} \tilde{b}_{\vec{k}} \rangle =
  \int \frac{d^{2} \vec{k} }{ (2 \pi)^{2} } v_{\vec{k}}^{2} $
   which is the quantum depletion of the condensate due to the dipole-dipole
   interaction. One can decompose the interaction Hamiltonian $
H_{int}$ in Eqn.\ref{first} into the coupling to the condensate part
$ H_{int}^{c}=\sum_{k_{z}}[ig(k_{z})\sqrt{N} a_{k_{z}}+h.c.] $ and to the quasi-particle part $
H_{int}^{q}=\sum_{k}[ig(k)a_{k}\tilde{b}_{\vec{k}}^{\dagger }+h.c.]
$. In the following, we study $ \vec{k}=0 $ and $ \vec{k} \neq 0 $ respectively.

The initial state of the exciton+ photons system in Fig.1 is taken to be:
\begin{equation}
|in\rangle = |BEC\rangle |0\rangle_{ph}
\label{in}
\end{equation}%

 Assuming a non-equilibrium steady exciton superfluid state has been reached, the output field
 $ a_{\vec{k}}^{out}(\omega )$ in the Fig.1 is  related to the input field by \cite{online,factori}:
\begin{eqnarray}
a_{\vec{k}}^{out}(\omega ) &=&[-1+ i \gamma _{\vec{k}}G_{n}(\vec{k},\omega +i%
\frac{\gamma _{k}}{2})]a_{\vec{k}}^{in}(\omega )  \nonumber \\
&& - i \gamma _{\vec{k}}G_{a}(\vec{k},\omega +i\frac{\gamma _{k}}{2})a_{-\vec{k}%
}^{in\dagger }(-\omega ),  \label{aout}
\end{eqnarray}%
where the normal and anomalous Green functions are \cite{online,factori}:
\begin{eqnarray}
G_{n}(\vec{k},\omega ) & =  & \frac{\omega +\epsilon _{\vec{k}}+\bar{n}V_{d}( \vec{k})}{\omega
^{2}-E^{2}(\vec{k})}   \nonumber   \\
G_{a}(\vec{k},\omega ) & =  &  - \frac{\bar{n}V_{d}(\vec{k})}{\omega
^{2}-E^{2}( \vec{k})}
\label{gnga}
\end{eqnarray}
where the $ \omega =\omega _{k}-\mu $. The exciton decay rate in the two Green functions are\cite{power}:
\begin{equation}
 \gamma _{\vec{k}}= 2 \pi D_{\vec{k}}(\mu )\left\vert g_{\vec{k}}(\omega
_{k}=\mu )\right\vert ^{2}
\label{gammak}
\end{equation}
which is independent of $ L_{z} $, so is an experimentally measurable quantity.  Just
from the rotational invariance, we can conclude that $ \gamma
_{\vec{k}} \sim const.+ |\vec{k}|^{2} $ as $ \vec{k} \rightarrow 0 $
as shown in Fig.3.

 From Ref.\cite{power}, one can also find the exciton operator \cite{tilde}:
\begin{eqnarray}
  b_{\vec{k}} (\omega ) &=& i \sqrt{\gamma _{\vec{k}}}[ G_{n}(\vec{k},\omega +i%
\frac{\gamma _{k}}{2}) a_{\vec{k}}^{in}(\omega )  \nonumber \\
& - & G_{a}(\vec{k},\omega +i\frac{\gamma _{k}}{2})a_{-\vec{k}%
}^{in\dagger }(-\omega ) ]  \label{b}
\end{eqnarray}%

\begin{figure}
\includegraphics[width=6cm]{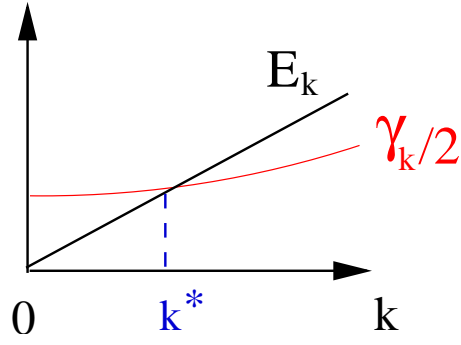}
\hspace{0.3cm}
\caption{ The energy spectrum versus the decay rate of the exciton
at a given in-plane momentum $\vec{k}$.  }
\label{fig3}
\end{figure}

  Note that the input-output formalism Eqns.\ref{aout},\ref{b} stands for a non-equilibrium dynamic process. It takes Eqn.\ref{in} as  the initial state,
  then solving the coupled Heisenberg equations of the photon and exciton operators. The steady state of the excitons has been reached,
  the photons are scattering from such an exciton steady state, so the Eqn.\ref{aout} can be considered as the scattering $ S $ matrix shown in the Fig.1.

  In the following sections, we will first use the Eqn.\ref{aout} to calculate the one photon normal and anomalous Green functions, then
  we will use the Eqn.\ref{b} to evaluate all the 5 normal and anomalous exciton correlation functions:
  the greater $ G^{>} $, lesser $ G^{<} $, retarded $ G^{R} $, advanced $ G^{A} $, the time-ordered $ G^{T} $ and the Keldysh component $ G^{K} $.

\section{ One photon normal and anomalous correlation functions }

  From Eqn.\ref{aout}, one can compute the photon umber spectrum\cite{power}:
\begin{eqnarray}
   S_{1}( \vec{k}, \omega ) & = & \langle  a^{out \dagger}_{\vec{k}} (\omega )
  a^{out}_{\vec{k}} (\omega ) \rangle_{in}  =    \gamma^{2} _{\vec{k}} | G_{a}(\vec{k},\omega +i\frac{\gamma _{k}}{2}) |^{2}
  \nonumber  \\
 & = & \frac{ \gamma^{2}_{\vec{k}} \bar{n}^{2}V_{d}^{2}(\vec{k})}{
\Omega ^{2}(\omega )+\gamma _{\vec{k}}^{2}E^{2}(\vec{k})}
\label{power}
\end{eqnarray}%
where $\Omega (\omega )=\omega
^{2}-E^{2}(\vec{k})+\gamma_{\vec{k}}^{2}/4$.

By using the Fourier transformation to $S_{1}(\vec{k}, \omega )$, one can get $%
G_{\pm }(\vec{k}, \tau )=G_{1}(\vec{k}, \tau )$
\begin{equation}
G_{1}(\vec{k}, \tau )=\frac{\bar{n}^{2}V_{d}^{2}(k)\gamma
_{k}}{4E(k)}[\frac{ e^{i(E(k)+i\frac{\gamma _{k}}{2})\tau
}}{E(k)+i\frac{\gamma _{k}}{2}}+h.c.] \label{g1}
\end{equation}
 which, obviously, only depends on the anomalous Green function $ G_{a}(\vec{k},\omega) $.

The normalized first order correlation function $g_{\pm }(\vec{k},
\tau )=g_{1}(\vec{k}, \tau ) =G_{1}(\vec{k}, \tau )/G_{1}(\vec{k},
\tau =0 ) $. When $ \tau > 0 $, it is given by:
\begin{equation}
g_{1}(\vec{k}, \tau )=e^{-\frac{\gamma _{\vec{k}}}{2}\tau }[\cos
(E(\vec{k})\tau )+ \frac{\gamma _{\vec{k}}}{2E(\vec{k})}\sin
(E(\vec{k})\tau )] \label{g1n}
\end{equation}
It turns out that the first order correlation function is
independent of the relation between $E(\vec{k})$ and $\gamma
_{\vec{k}}/2$.

 Similarly, one can find:
\begin{equation}
  \langle  a^{out }_{\vec{k}}(\omega )
  a^{out \dagger}_{\vec{k}} (\omega ) \rangle=|-1 + \gamma _{\vec{k}} G_{n}(\vec{k},\omega +i\frac{\gamma _{k}}{2})
  |^{2}
\label{anti}
\end{equation}
  which, in sharp contrast to Eqn.\ref{g1}, only depends on the normal Green function $ G_{n}(\vec{k},\omega) $.

  The power spectrum experiments \cite{power} only measure the normal ordered
  one photon correlation function Eqn.\ref{g1}. The anti-normal ordered
  one photon correlation function Eqn.\ref{anti} maybe measured
  by a transmission spectrum with a weak external pumping.

  One can also calculate the  anomalous one photon Green function:
\begin{eqnarray}
   \langle  a^{out }_{\vec{k}}(\omega )
  a^{out }_{-\vec{k}} (-\omega ) \rangle & = & [-1 + \gamma _{\vec{k}} G_{n}(\vec{k},\omega +i\frac{\gamma_{k}}{2})]
                  \nonumber  \\
  & \times & \gamma _{\vec{k}} G_{a}(\vec{k},-\omega + i\frac{\gamma_{k}}{2})
\label{aa}
\end{eqnarray}
   which  depends on both the normal and the anomalous Green functions.

  By changing $ \vec{k} \rightarrow -\vec{k}, \omega
\rightarrow -\omega $ in Eqn.\ref{aa}, one can get:
\begin{eqnarray}
   \langle  a^{out }_{-\vec{k}}(-\omega )
  a^{out }_{\vec{k}} (\omega ) \rangle & = & [-1 + \gamma _{\vec{k}} G_{n}(\vec{k},-\omega +i\frac{\gamma_{k}}{2})]
                  \nonumber  \\
  & \times & \gamma _{\vec{k}} G_{a}(\vec{k},\omega + i\frac{\gamma_{k}}{2})
\label{aam}
\end{eqnarray}

   From the explicit forms of $ G_n, G_a $ listed in Eqn.\ref{gnga}, one can observe that the right hand side of Eqn.\ref{aa} is a even
   function of $ \omega $, so:
\begin{equation}
   \langle  a^{out }_{\vec{k}}(\omega )
  a^{out }_{-\vec{k}} (-\omega ) \rangle  =   \langle  a^{out }_{-\vec{k}}(-\omega )
  a^{out }_{\vec{k}} (\omega ) \rangle
\label{aaequal}
\end{equation}

    This identity is expected because both the input and output fields obey the Bose commutation relations \cite{power}:
\begin{eqnarray}
& & [a_{\vec{k}}^{in}(t),a_{ \vec{k}^{\prime} }^{in \dagger
}(t^{\prime})]=[a_{ \vec{k}}^{out}(t),a_{ \vec{k}^{\prime}
}^{out\dagger }(t^{\prime})]=\delta_{\vec{k},\vec{k}^{\prime}
}\delta (t-t^{\prime})   \nonumber   \\
& & [a_{\vec{k}}^{in}(t),a_{\vec{k}^{\prime}}^{in}(t^{\prime})]
=[a_{\vec{k}}^{out}(t),a_{\vec{k}^{\prime}}^{out}(t^{\prime})]= 0
\label{comm}
\end{eqnarray}
    so that  $  \langle a_{ \vec{k}}^{out}(t) a_{ -\vec{k} }^{out}(0) \rangle_{in}
    =  \langle a_{ -\vec{k} }^{out}(0) a_{ \vec{k}}^{out}(t) \rangle_{in} $ whose Fourier transform leads to Eqn.\ref{aaequal}.

    One can also find the Fourier transforms of Eqn.\ref{aa} and \ref{aam}:
\begin{eqnarray}
    F_{1}(\tau) & = & \int \frac{ d \omega }{ 2 \pi} e^{ i \omega
   \tau }  \langle  a^{out }_{-\vec{k}}(-\omega )
    a^{out }_{\vec{k}} (\omega ) \rangle   \nonumber   \\
    F_{2}(\tau) & =  & \int \frac{ d \omega }{ 2 \pi} e^{ i \omega
   \tau }  \langle  a^{out }_{\vec{k}}(\omega )
    a^{out }_{-\vec{k}} (-\omega ) \rangle
\label{f1f2}
\end{eqnarray}
    Using Eqn.\ref{aaequal}, one can see that $ F_{1}(\tau)=
    F_{2}(\tau) $. Then the normalized anomalous one photon correlation function
    in the time domain is given by:
\begin{eqnarray}
f_{1}(\tau ) &=&  F_{1}(\tau)/ G_{1}(0) =
-\frac{E^{2}(\vec{k})+\frac{\gamma
_{\vec{k}}^{2}}{4}}{\bar{n}V_{d}(\vec{k})}e^{-\frac{\gamma _{\vec{k}}}{2}\tau }  \nonumber \\
& \times &[u_{\vec{k}}^{2}\frac{e^{-iE(\vec{k})\tau
}}{E(\vec{k})-i\frac{\gamma
_{\vec{k}}}{2}}+v_{\vec{k}}^{2}\frac{e^{iE(\vec{k})\tau
}}{E(\vec{k})+i\frac{ \gamma _{\vec{k}}}{2}}] \label{f1n}
\end{eqnarray}
   which was used in calculating the two photon correlation functions in \cite{squ}.

   Measuring these photon anomalous Green functions are through the
   phase sensitive homodyne measurements discussed in the \cite{squ}.

\section{ One exciton  normal correlation functions  }

 From Eqn.\ref{b}, one can compute the normal exciton correlation function:
\begin{eqnarray}
 i G^{<}_{n}( \vec{k}, \omega ) & = & \langle  b^{\dagger}_{\vec{k}} (\omega )
  b_{\vec{k}} (\omega ) \rangle_{in}  =    \gamma _{\vec{k}} | G_{a}(\vec{k},\omega +i\frac{\gamma _{k}}{2}) |^{2}
  \nonumber  \\
 & = & \frac{ \gamma _{\vec{k}} \bar{n}^{2}V_{d}^{2}(\vec{k})}{
\Omega ^{2}(\omega )+\gamma _{\vec{k}}^{2}E^{2}(\vec{k})}
\label{gln}
\end{eqnarray}%
where $\Omega (\omega )=\omega
^{2}-E^{2}(\vec{k})+\gamma_{\vec{k}}^{2}/4$.
Note that it only depends on the anomalous Green function.
  Obviously, the normalized first order exciton correlation function
\begin{equation}
  g_{1b}(\vec{k}, \tau ) =G_{1b}(\vec{k}, \tau )/G_{1b}(\vec{k}, \tau =0 )= g_{1}(\vec{k}, \tau )
\label{g1b}
\end{equation}
 is also given by Eqn.\ref{g1n} and shown in Fig.4.

\begin{figure}
\includegraphics[width=7cm]{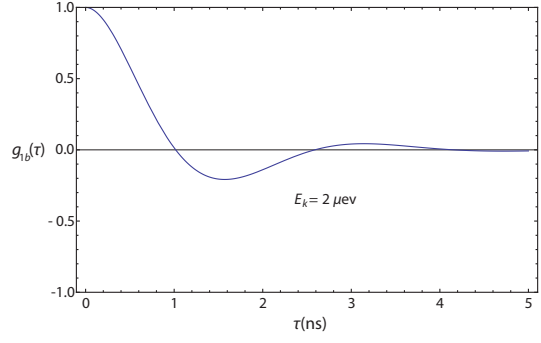}
\caption{ The one exciton correlation function of Eqn.\ref{g1b} at the experimental relevant values \cite{diff}
 $ n V_d(\vec{k}) =5 \mu e V, \gamma_{\vec{k}}/2= 1 \mu e V $
 at the quasi-particle energies $ E( \vec{k} ) =2 \mu e V $. The same values of the $ n V_d(\vec{k}), \gamma_{\vec{k}}/2 $
 will also be used \cite{draw} in Fig.5-10. }
\label{fig4}
\end{figure}

 From Eqn.\ref{b}, one can compute the exciton correlation function:
\begin{eqnarray}
 i G^{>}_{n}( \vec{k}, \omega ) & = & \langle  b_{\vec{k}} (\omega )
  b^{\dagger}_{\vec{k}} (\omega ) \rangle_{in}  =  \gamma _{\vec{k}} | G_{n}(\vec{k},\omega +i\frac{\gamma _{k}}{2}) |^{2}
  \nonumber  \\
 & = & \frac{ \gamma _{\vec{k}}[ ( \omega + \epsilon_{\vec{k}} + \bar{n} V_{d}(\vec{k}))^{2} + (\gamma _{\vec{k}}/2)^{2} ] }{
\Omega ^{2}(\omega )+\gamma _{\vec{k}}^{2}E^{2}(\vec{k})}
\label{ggn}
\end{eqnarray}%
which only depends on the
normal Green function, in sharp contrast to Eqn.\ref{gln}.

 From Eqn.\ref{ggn} and Eqn.\ref{gln}, one can find the normal spectral  weight
\begin{eqnarray}
 \rho_{n}( \vec{k}, \omega )  =  \langle  b_{\vec{k}} (\omega )  b^{\dagger}_{\vec{k}} (\omega )
 \rangle_{in}-\langle  b^{\dagger}_{\vec{k}} (\omega )  b_{\vec{k}} (\omega )
 \rangle_{in}  \nonumber  \\
   = i [ G^{>}_{n}( \vec{k}, \omega )-G^{<}_{n}( \vec{k}, \omega )] ~~~~~~~~~~~~  \nonumber \\
   =   \gamma _{\vec{k}} [  | G_{n}(\vec{k},\omega +i\frac{\gamma _{k}}{2}) |^{2}
   - | G_{a}(\vec{k},\omega +i\frac{\gamma _{k}}{2}) |^{2}]
\label{nw}
\end{eqnarray}

   Using the Eqn.\ref{gnga}, one can manipulates the last equation in \ref{nw} into a surprisingly simple result:
\begin{equation}
   \rho_{n}( \vec{k}, \omega ) = i [ G_{n}(\vec{k},\omega + i\frac{\gamma _{k}}{2})-  G_{n}(\vec{k},\omega -i\frac{\gamma _{k}}{2})]
\label{nw1}
\end{equation}

   One can also calculate the Retarded and Advanced normal Green function:
\begin{eqnarray}
 i G^{R}_{n}(\vec{k}, t ) & = & \theta(t)[ b_{\vec{k}}(t), b^{\dagger}_{\vec{k}}(0) ]
  \nonumber  \\
 i G^{A}_{n}(\vec{k}, t )  & =  & -\theta(-t)[ b_{\vec{k}}(t), b^{\dagger}_{\vec{k}}(0) ]
\label{ran}
\end{eqnarray}
   Their Fourier transforms lead to their corresponding normal spectral weight representations:
\begin{eqnarray}
  G^{R}_{n}(\vec{k},\omega ) = \int \frac{ d \omega^{\prime} }{ 2 \pi } \frac{ \rho_{n}( \vec{k}, \omega^{\prime} ) }{ \omega - \omega^{\prime} +i \eta}
  \nonumber  \\
  G^{A}_{n}(\vec{k},\omega ) = \int \frac{ d \omega^{\prime} }{ 2 \pi } \frac{ \rho_{n}( \vec{k}, \omega^{\prime} ) }{ \omega - \omega^{\prime} -i \eta}
\label{ran1}
\end{eqnarray}
   Plugging the Eqn.\ref{nw1} into the above equations leads to:
\begin{eqnarray}
  G^{R}_{n}(\vec{k},\omega ) = G_{n}(\vec{k},\omega + i\frac{\gamma _{k}}{2})
  \nonumber  \\
  G^{A}_{n}(\vec{k},\omega ) = G_{n}(\vec{k},\omega -i\frac{\gamma _{k}}{2})
\label{ran2}
\end{eqnarray}
   whose physical pictures are clear: the Retarded or Advanced normal Green functions can be achieved simply
   by just putting the decay rate $  \pm \frac{\gamma _{k}}{2} $ in the non-dissipative ones in Eqn.\ref{gnga}.

  From Eqns.\ref{nw},\ref{nw1},\ref{ran2}, we can check the identity:
\begin{equation}
  G^{R}_{n}- G^{A}_{n}= G^{>}_{n}-G^{<}_{n}
\end{equation}
   as expected.

  We can also compute the Time ordered exciton Green  function
\begin{eqnarray}
   i G^{T}_{n}( \vec{k}, t )  = \langle T b_{\vec{k}}(t) b^{\dagger}_{\vec{k}}(0)\rangle   ~~~~~~~~~
                  \nonumber  \\
     =  \theta(t) \langle  b_{\vec{k}}(t) b^{\dagger}_{\vec{k}}(0)\rangle + \theta(-t)  \langle  b^{\dagger}_{\vec{k}}(0)  b_{\vec{k}}(t) \rangle
\end{eqnarray}
   Its Fourier transform leads to:
\begin{equation}
  G^{T}_{n}( \vec{k}, \omega )= \int \frac{ d \omega^{\prime}}{2 \pi i } [
  \frac{  G^{>}_{n}(\vec{k},\omega^{\prime} ) }{ \omega^{\prime}-
  \omega - i \eta } -  \frac{  G^{<}_{n}(\vec{k},\omega^{\prime} ) }{ \omega^{\prime}-
  \omega + i \eta } ]
\label{gtn}
\end{equation}
    Substituting Eqn.\ref{gln}, \ref{ggn} into the above equation and paying special attentions to the pole structures in
    the Eqn.\ref{gtn}, one can show  that
\begin{eqnarray}
  G^{T}_{n}( \vec{k}, \omega ) & = & [1 -i \gamma _{\vec{k}} G_{n}(\vec{k},\omega +i\frac{\gamma_{k}}{2})] G_{n}(\vec{k},\omega - i\frac{\gamma_{k}}{2})
                        \nonumber  \\
   & = &  [1 -i \gamma _{\vec{k}} G^{R}_{n}(\vec{k},\omega )] G^{A}_{n}(\vec{k},\omega )
\label{gtn1}
\end{eqnarray}

   Finally,  from Eqn.\ref{ggn} and Eqn.\ref{gln}, we can easily get the Keldysh component of the Green function:
\begin{eqnarray}
  i  G^{K}_{n}( \vec{k}, \omega ) = i [ G^{>}_{n}( \vec{k}, \omega )+ G^{<}_{n}( \vec{k}, \omega )] ~~~~~~~~~~~~  \nonumber \\
   = \frac{ \gamma _{\vec{k}}[ ( \omega + \epsilon_{\vec{k}} + \bar{n} V_{d}(\vec{k}))^{2}
   + (\bar{n} V_{d}(\vec{k}))^{2} + (\gamma _{\vec{k}}/2)^{2} ] }{
\Omega ^{2}(\omega )+\gamma _{\vec{k}}^{2}E^{2}(\vec{k})}
\label{keln}
\end{eqnarray}
   which is shown in the Fig.6. It is neither a even nor an odd function.

   It is well known that for an equilibrium system, the  $ G^{>}_{n} $ and $ G^{<}_{n} $  are related by the Fluctuation-Dissipation Theorem (FDT)
   instead of being independent. At $ T=0 $, the FDT takes the form
\begin{equation}
  i [ G^{>}_{n}( \vec{k}, \omega ) + G^{<}_{n}( \vec{k}, \omega )] =i sgn \omega  [ G^{>}_{n}( \vec{k}, \omega )-G^{<}_{n}( \vec{k}, \omega )]
\label{fdtn}
\end{equation}
   However, for a non-equilibrium system with the initial input state Eqn.\ref{in},
   the $ G^{>}_{n} $ in Eqn.\ref{ggn} and the $ G^{<}_{n} $ Eqn.\ref{gln} {\sl violate } the FDT Eqn.\ref{fdtn}.
   Eqn.\ref{ran1} and \ref{gtn} lead to $ G^{R}_{n} $, $ G^{A}_{n} $ and $ G^{T}_{n} $ respectively.
   This is an salient feature expected for any non-equilibrium system. The initial input state Eqn.\ref{in} is not even an
   eigen-state, let alone the ground state of the total Hamiltonian Eqn.\ref{first}.
   It is this fact which leads to the dynamic photon scattering process encoded in
   Eqn.\ref{aout} and \ref{b}.
   The spectral weight $ \rho_n( \vec{k}, \omega ) $ in Eqn.\ref{nw} and the Keldysh component
   $ G^{K}_{n}( \vec{k}, \omega ) $ in  Eqn.\ref{keln} automatically follow.
   Detailed comparisons with the corresponding equilibrium system will be given in Sect.IX and  the conclusion section.

\section{ One exciton anomalous   correlation functions  }

From Eqn.\ref{b}, one can also compute the abnormal exciton
correlation function $ i G^{>}_{a}(\vec{k},\omega )= \langle
b_{\vec{k}} (\omega ) b_{-\vec{k}} (-\omega ) \rangle $:
\begin{eqnarray}
 i G^{>}_{a}(\vec{k},\omega )=  \gamma _{\vec{k}} G_{n}(\vec{k},\omega +i\frac{\gamma _{k}}{2}) G_{a}(\vec{k},-\omega +i\frac{\gamma _{k}}{2})
  \nonumber  \\
  =  - \frac{ \gamma _{\vec{k}}(  \omega +i\frac{\gamma _{k}}{2} + \epsilon_{\vec{k}} + \bar{n} V_{d}(\vec{k})) \bar{n} V_{d}(\vec{k}) }{
\Omega ^{2}(\omega )+\gamma _{\vec{k}}^{2}E^{2}(\vec{k})} ~~~~~~~~~~
\label{gga}
\end{eqnarray}
which depends on both the normal and the anomalous Green function.

   By changing $ \vec{k} \rightarrow -\vec{k}, \omega
\rightarrow -\omega $ in Eqn.\ref{gga}, one can get $
 i G^{<}_{a}(\vec{k},\omega )= \langle b_{-\vec{k}} (-\omega )
b_{\vec{k}} (\omega ) \rangle_{in} $:
\begin{eqnarray}
i G^{<}_{a}(\vec{k},\omega )=  \gamma _{\vec{k}} G_{n}(\vec{k},-\omega +i\frac{\gamma _{k}}{2}) G_{a}(\vec{k}, \omega +i\frac{\gamma _{k}}{2})
\label{gla}
\end{eqnarray}

From Eqn.\ref{gga} and \ref{gla}, one find the  abnormal spectral
weight
\begin{eqnarray}
 \rho_{a}( \vec{k}, \omega )  = i [ G^{>}_{a}(\vec{k},\omega ) - G^{<}_{a}(\vec{k},\omega )  ]
\end{eqnarray}
   When using  Eqn.\ref{aa}, \ref{aam} and \ref{aaequal}, one can manipulate it into a surprisingly simple result:
\begin{eqnarray}
 \rho_{a}( \vec{k}, \omega )  = i [ G^{>}_{a}(\vec{k},\omega ) - G^{<}_{a}(\vec{k},\omega )  ]  ~~~~\nonumber  \\
   = i [   G_{a}(\vec{k},\omega +i\frac{\gamma _{k}}{2})-G_{a}(\vec{k},\omega - i\frac{\gamma _{k}}{2}) ]
\label{aw}
\end{eqnarray}

  In contrast to Eqn.\ref{aaequal} for photons,  $ G^{>}_{a}(\vec{k},\omega ) \neq G^{<}_{a}(\vec{k},\omega
  ) $, so $  \rho_{a}( \vec{k}, \omega ) \neq 0 $. This is expected, because $ \langle
   b_{\vec{k}} ( t ) b_{-\vec{k}} ( 0 ) \rangle \neq \langle
   b_{-\vec{k}} ( 0 ) b_{\vec{k}} ( t ) \rangle $.

   One can also find the Fourier transforms of Eqn.\ref{gla} and \ref{gga}:
\begin{eqnarray}
    F_{1b}(\tau) & = & \int \frac{ d \omega }{ 2 \pi} e^{ i \omega
   \tau }  i G^{<}_{a}(\vec{k},\omega )  \nonumber   \\
    F_{2b}(\tau) & =  & \int \frac{ d \omega }{ 2 \pi} e^{ i \omega
   \tau }  i G^{>}_{a}(\vec{k},\omega )
\label{f1f2bbig}
\end{eqnarray}
    In contrast to the corresponding quantities for photons $ F_{1}(\tau)=
    F_{2}(\tau) $ listed in Eqn.\ref{f1f2}, $ F_{1b}(\tau) \neq  F_{2b}(\tau) $.
    In fact, one can see that $  F_{1b}(\tau)= F_{1}(\tau)/\gamma_{\vec{k}},
    F_{2b}(\tau)= F_{1b}(\tau)-i \bar{n} V_{d}(k) e^{-\frac{\gamma_{\vec{k}}}{2}\tau } \frac{ \sin(E(\vec{k})\tau )}{E(\vec{k}) }  $.

    Then the normalized anomalous one exciton correlation function
    in the time domain is given by:
\begin{eqnarray}
f_{1b}(\tau ) &=&  F_{1b}(\tau)/ G_{1b}(0) = f_{1}(\tau )
\nonumber \\
f_{2b}(\tau ) &=&  F_{2b}(\tau)/ G_{1b}(0) = f_{1b}(\tau )-i 2
\frac{E^{2}(\vec{k})+\frac{\gamma
_{\vec{k}}^{2}}{4}}{\bar{n}V_{d}(\vec{k})} \nonumber  \\
 & \times & e^{-\frac{\gamma
_{\vec{k}}}{2}\tau }  \frac{ \sin(E(\vec{k})\tau )}{E(\vec{k}) }
\label{f1f2b}
\end{eqnarray}
     which will be useful in calculating two exciton correlation
     functions in the Sect.VIII.

 One can also calculate the Retarded and Advanced abnormal Green function:
\begin{eqnarray}
 i G^{R}_{a}(\vec{k}, t ) & = & \theta(t)[ b_{\vec{k}}(t), b_{-\vec{k}}(0) ]
  \nonumber  \\
 i G^{A}_{a}(\vec{k}, t )  & =  & -\theta(-t)[ b_{\vec{k}}(t), b_{-\vec{k}}(0) ]
\label{raa}
\end{eqnarray}
   Their Fourier transforms lead to their corresponding spectral weight representations:
\begin{eqnarray}
  G^{R}_{a}(\vec{k},\omega ) = \int \frac{ d \omega^{\prime} }{ 2 \pi } \frac{ \rho_{a}( \vec{k}, \omega^{\prime} ) }{ \omega - \omega^{\prime} +i \eta}
  \nonumber  \\
  G^{A}_{a}(\vec{k},\omega ) = \int \frac{ d \omega^{\prime} }{ 2 \pi } \frac{ \rho_{a}( \vec{k}, \omega^{\prime} ) }{ \omega - \omega^{\prime} -i \eta}
\label{raa1}
\end{eqnarray}
   Plugging the Eqn.\ref{aw} into the above equation leads to:
\begin{eqnarray}
  G^{R}_{a}(\vec{k},\omega ) = G_{a}(\vec{k},\omega + i\frac{\gamma _{k}}{2})
  \nonumber  \\
  G^{A}_{a}(\vec{k},\omega ) = G_{a}(\vec{k},\omega -i\frac{\gamma _{k}}{2})
\label{raa2}
\end{eqnarray}
   whose physical pictures are clear: the Retarded or Advanced anomalous Green functions can be achieved simply
   by just putting the decay rate $  \pm \frac{\gamma _{k}}{2} $ in the non-dissipative ones in Eqn.\ref{gnga}.

  From Eqns.\ref{aw}, we can check the identity:
\begin{equation}
  G^{R}_{a}- G^{A}_{a}= G^{>}_{a}-G^{<}_{a}
\end{equation}
   as expected.

  We can compute the Time ordered anomalous exciton Green function:
\begin{eqnarray}
  i G^{T}_{a}( \vec{k}, t )  =  \langle T b_{\vec{k}}(t)  b_{-\vec{k}}(0)\rangle    ~~~~~~~~~         \nonumber  \\
   =  \theta(t)  \langle  b_{\vec{k}}(t)
  b_{-\vec{k}}(0)\rangle + \theta(-t)  \langle  b_{-\vec{k}}(0)  b_{\vec{k}}(t) \rangle
\end{eqnarray}

  Its Fourier transform lead to:
\begin{equation}
  G^{T}_{a}( \vec{k}, \omega )= \int \frac{ d \omega^{\prime}}{2 \pi i } [
  \frac{  G^{>}_{a}(\vec{k},\omega^{\prime} ) }{ \omega^{\prime}-
  \omega - i \eta } -  \frac{  G^{<}_{a}(\vec{k},\omega^{\prime} ) }{ \omega^{\prime}-
  \omega + i \eta } ]
\label{gta}
\end{equation}
    Substituting Eqn.\ref{gga},\ref{gla} into the above equation and paying special attentions to the pole structures in
    the Eqn.\ref{gta}, one can show  that
\begin{eqnarray}
  G^{T}_{a}( \vec{k}, \omega ) & = & [1 -i \gamma _{\vec{k}} G_{n}(\vec{k},\omega +i\frac{\gamma_{k}}{2})]
  G_{a}(\vec{k},\omega - i\frac{\gamma_{k}}{2})   \nonumber  \\
     & = &  G^{A}_{a}(\vec{k},\omega )  +  G^{>}_{a}(\vec{k},\omega )   \nonumber  \\
     & = &  G^{R}_{a}(\vec{k},\omega )  +  G^{<}_{a}(\vec{k},\omega )
\label{gta1}
\end{eqnarray}

 Finally,  from Eqn.\ref{ggn} and Eqn.\ref{gln}, we can easily get the Keldysh component of the anomalous Green function:
\begin{eqnarray}
  i  G^{K}_{a}( \vec{k}, \omega ) = i [ G^{>}_{a}( \vec{k}, \omega )+ G^{<}_{a}( \vec{k}, \omega )]  \nonumber \\
   =  -  \frac{ 2 \gamma _{\vec{k}}( i\frac{\gamma _{k}}{2} + \epsilon_{\vec{k}} + \bar{n} V_{d}(\vec{k})) \bar{n} V_{d}(\vec{k}) }{
\Omega ^{2}(\omega )+\gamma _{\vec{k}}^{2}E^{2}(\vec{k})}
\label{kela}
\end{eqnarray}
   which is a complex even function. Its real and imaginary part are shown in the Fig.8 and Fig.9 respectively.

   Similar to the normal exciton correlation functions discussed in the previous section,  as expected for an
   non-equilibrium system, the $ G^{>}_{a} $ in Eqn.\ref{gga} and $ G^{<}_{a} $ Eqn.\ref{gla} are two independent Green functions,
   so violates the FDT for anomalous Green function:
\begin{equation}
  i [ G^{>}_{a}( \vec{k}, \omega ) + G^{<}_{a}( \vec{k}, \omega )] =i sgn \omega  [ G^{>}_{a}( \vec{k}, \omega )-G^{<}_{a}( \vec{k}, \omega )]
\label{fdta}
\end{equation}
   Eqn.\ref{raa1} and \ref{gta} lead to $ G^{R}_{a} $, $ G^{A}_{a} $ and $ G^{T}_{a} $ respectively.
   The spectral weight $ \rho_n( \vec{k}, \omega ) $ in Eqn.\ref{aw} and the Keldysh component
   $ G^{K}_{n}( \vec{k}, \omega ) $ in  Eqn.\ref{kela} automatically follow.
   Detailed comparisons with the corresponding equilibrium system will be given in Sect.IX and  the conclusion section.


\section{ Input-output relations between photons and excitons }

 From the exciton condensation Eqn.\ref{conden} at $ \vec{k} = 0 $ and the emitted photon number at $ \vec{k} = 0 $ Eqn.4 in Ref.\ref{power}
 ( which is the same as the photon condensation  Eqn.\ref{photoncon} ),
 one can immediately see the relation between the two:
\begin{equation}
  N_{ph}= N_{b} \gamma_0
\label{inoutcon}
\end{equation}
    where the $ \gamma_0 $ is the exciton decay rate at $ \vec{k}=0 $. This relation relates the condensation
    of the emitted photon to that of excitons.

 From Eqn.\ref{power},\ref{gln}, one can see that
\begin{equation}
   \langle  a^{out \dagger}_{\vec{k}}(\omega )
  a^{out}_{\vec{k}} (\omega ) \rangle_{in} = \gamma _{\vec{k}} \langle  b^{\dagger}_{\vec{k}} (\omega )
  b_{\vec{k}} (\omega ) \rangle_{in}
\label{inout}
\end{equation}
  which relates the emitted photon spectrum to the internal
  normal ordered exciton correlations. A similar relation was discussed in the context of optical cavities \cite{book1,book2}.
  Here it is in the spontaneous photon emission from an exciton superfluid in the absence of any cavities.
  This relation shows that the angle resolved photon power spectrum studied in \cite{power} can reflect precisely the normal ordered normal
  exciton correlation function $ i G^{>}_{n} $.

  However, one can also see that due to the extra first term $ 1 $ in Eqn.\ref{anti}, there is no such simple
  input-out relation between  $ \langle  a^{out }_{\vec{k}}(\omega )
  a^{out \dagger}_{\vec{k}} (\omega ) \rangle $ and  $ \langle  b_{\vec{k}} (\omega )
  b^{\dagger}_{\vec{k}} (\omega ) \rangle $ listed in Eqn.\ref{ggn}.

  Now we will try to explore the input-output relation between the
  anomalous photon correlation function in Eqn.\ref{aa} and that of the excitons in
  Eqn.\ref{gga}. Obviously, they are not simply related as the Eqn.\ref{inout}. However,
  by comparing Eqn.\ref{aa} with Eqn.\ref{gta1}, one can establish  the input-output relation between the
  anomalous photon correlation function with the time-ordered correlation function of the  excitons:
\begin{equation}
 \langle  a^{out }_{\vec{k}}(\omega )
  a^{out }_{-\vec{k}} (-\omega ) \rangle=  \gamma _{\vec{k}}  i G^{T}_{a}( \vec{k}, \omega )
\label{inoutt}
\end{equation}
  which relates the emitted photon  two mode squeezing spectrum to the internal
  time-ordered exciton correlations.  A similar relation was discussed in the context of optical cavities \cite{book1,book2}.
  Here it is in the spontaneous photon emission from an exciton superfluid in the absence of any cavities.
  This relation shows that the phase sensitive homedyne measurements  to measure the two mode squeezing spectrum \cite{squ} can
  reflect precisely the Time ordered anomalous   exciton correlation function $ i G^{T}_{a} $.
  It is instructive to compare  Eqn.\ref{inoutt} which involves the Time-ordered with Eqn.\ref{inout} which involves just normal ordered.


\section{ The quasi-particle spectra and distribution functions in a non-equilibrium stationary
exciton superfluid }

  There are three kinds of Photoluminescence measurements. One kind is the angle resolved  power spectrum (ARPS) shown in Fig.3 in \cite{power}.
  The second kind is the  two mode squeezing spectrum shown in Fig.1b-3a in \cite{squ}.
  In any non-equilibrium system, one need both the spectral weight
  $   \rho_{n/a}( \vec{k}, \omega ) = i [ G^{>}_{n/a}( \vec{k}, \omega ) -  G^{<}_{n/a}( \vec{k}, \omega )] $  and the Keldysh Green function component
  $   G^{K}_{n/a}( \vec{k}, \omega ) =  G^{>}_{n/a}( \vec{k}, \omega )+ G^{<}_{n/a}( \vec{k}, \omega ) $ to characterize the spectral weight
  and the distribution function respectively.   In this section, we will explore the relations between the two kinds of PL
  with the exciton spectral weights and distribution functions.
  Due to the lack of FDT in the non-equilibrium exciton superfluid system, these relations are non-trivial, so need to be studied in details.
  The third kind is the two photon correlations measurement shown in
  Fig.3b in \cite{squ}. In the next section, we will explore its relation with the two exciton correlation functions.

\subsection{ Connections among the ARPS, the exciton normal spectral weight and distribution function }

\begin{figure}
\includegraphics[width=8cm]{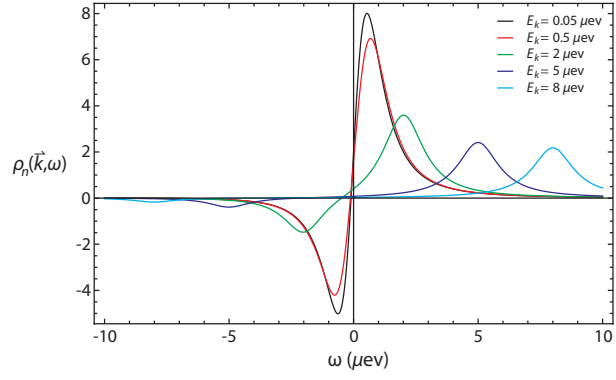}
\hspace{0.3cm}
\caption{ The normal spectral weight Eqn.\ref{lorn} at different quasi-particle energies
at $ E( \vec{k} ) < \gamma_{\vec{k}}/2 $: $ E( \vec{k} ) = 0.05, 0.5 $ and
at $ E( \vec{k} ) > \gamma_{\vec{k}}/2 $: $ E( \vec{k} ) = 2, 5, 8 \mu e V $. }
\label{fig5}
\end{figure}

  From Eqn.\ref{nw1}, we can get the normal spectral weight $  \rho_{n}( \vec{k}, \omega )   =  -2 Im G^{R}_{n}( \vec{k}, \omega ) $:
\begin{eqnarray}
 \rho_{n}( \vec{k}, \omega )  & = & u^{2}_{\vec{k}} \frac{ \gamma_{\vec{k}} }{ (\omega-E(\vec{k} ))^{2}+ (\gamma_{\vec{k}}/2)^{2} }
                    \nonumber  \\
     & - & v^{2}_{\vec{k}} \frac{ \gamma_{\vec{k}} }{ (\omega + E(\vec{k} ))^{2}+ (\gamma_{\vec{k}}/2)^{2} }
\label{lorn}
\end{eqnarray}
   where $ u^{2}_{\vec{k}}-v^{2}_{\vec{k}}= 1 $.  Obviously, $ \rho_{n}( \vec{k}, \omega )  $ satisfies the sum rule:
\begin{equation}
     \int \frac{ d \omega }{ 2 \pi}   \rho_{n}( \vec{k}, \omega ) =  u^{2}_{\vec{k}}-v^{2}_{\vec{k}}= 1
\end{equation}
     as expected from its definition.

   It is shown in Fig.5 for several different quasi-particle energies $ E(\vec{k}) $.
   At $ E(\vec{k} ) =0 $, the $ \rho_{n}( \vec{k}=0, \omega ) = \frac{ \gamma_{\vec{k}} }{ \omega^{2}+ (\gamma_{\vec{k}}/2)^{2} } > 0 $  is a single
   Lorentizan centered at $ \omega =0  $ with width $ \gamma_{\vec{k}}/2 $. At $ E(\vec{k} )  > 0 $,
   it  consists of two Lorentizan centered at $ \pm E(\vec{k} ) $ with width $ \gamma_{\vec{k}}/2 $
   with the spectral weight $ u^{2}_{\vec{k}} $ and $ -v^{2}_{\vec{k}} $ respectively.
   Only when $ E(\vec{k} ) \gg \gamma_{\vec{k}}/2 $ in Fig.5, namely, when $ k \gg k^{*} $ in the Fig.3, the two Lorentizans are well separated.
   Note that $ \rho_{n}( \vec{k}, \omega=0 ) = \frac{ \gamma_{\vec{k}} }{ E^{2}(\vec{k})+ (\gamma_{\vec{k}}/2)^{2} } > 0 $
   is large when $ E(\vec{k} ) \leq \gamma_{\vec{k}}/2 $.
   The positive value at $ \omega=0 $  can be seen in Fig.5 when $ E(\vec{k} ) \leq \gamma_{\vec{k}}/2 $.
   The normal distribution function $ i G^{K}_{n} $ is shown in Fig.6.

\begin{figure}
\includegraphics[width=8cm]{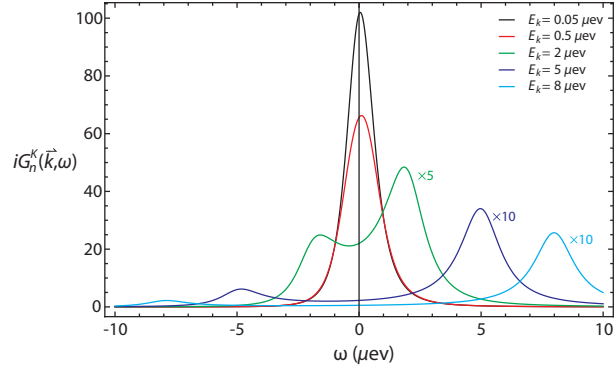}
\hspace{0.3cm}
\caption{ The normal Keldysh component $ i G^{K}_{n} $  at different quasi-particle energies $ E( \vec{k} ) = 0.05, 0.5, 2, 5, 8 \mu e V $. }
\label{fig6}
\end{figure}

  What the first class of  PL experiment measured is the angle resolved  power spectrum (ARPS) shown in Fig.3 in \cite{power}. The excitation spectrum
  and the distribution function of the exciton superfluid system to be probed by transport measurements is given by  Eqn.\ref{lorn} and \ref{keln} respectively.
  So it is important to explore the relation
  between what is measured by the PL and the system's properties to be probed by transport measurements.
  Such a relation is given by the input-output relation Eqn.\ref{inout},  Eqn.\ref{gln} and Eqn.\ref{lorn}.
  One can see qualitatively the same physics with slightly quantitatively different numbers.
  When $ k < k^{*} $, the Fig.5 shows that the quasi-particle is not well
  defined. Correspondingly, the ARPS in the Fig.3 in \cite{power} has just one peak  pinned at $ \omega_{k}= \mu $ with the
  width $ \gamma_{k} $. The MDC has large values at $ k < k^{*} $.
  The EDC has large values at $ \omega < \gamma_{k} $. From the Fig.6, one can see that the distribution function  has just one peak  pinned at $ \omega_{k}= \mu $ with the  width $ \gamma_{k} $.
  When $ k \gg k^{*} $, the Fig.5 shows that  the two Lorentizans centered at $ \pm E(\vec{k} ) $ with the width $ \gamma_{\vec{k}}/2 $ are well separated
  with two different spectral weights  $ u^{2}_{\vec{k}} $ and $ -v^{2}_{\vec{k}} $ respectively. The quasi-particles at  $ \pm E(\vec{k} ) $ are well
  defined. Correspondingly, the ARPS in the Fig.1 in \cite{power} has two well defined symmetric quasi-particles peaks at
  $ \omega_{k}=  \mu \pm \lbrack E^{2}(\vec{k})-\gamma _{\vec{k}}^{2}/4]^{1/2}$
  with the width $ \gamma_{k}/2 $. From the Fig.6, one can see that the distribution function split into two asymmetric peaks at
  $ \omega_{k}=  \mu \pm \lbrack E^{2}(\vec{k})-\gamma _{\vec{k}}^{2}/4]^{1/2}$  with the width $ \gamma_{k}/2 $.
  It is instructive to observe some differences between
  the $ \rho_{n}( \vec{k}, \omega )$ shown in the Fig.5 and the ARPS shown in the Fig.3 in \cite{power}: (1) There are two symmetry peaks in the latter,
  but the two peaks  in the former are  as-symmetric with two different spectral weights  $ u^{2}_{\vec{k}} $ and $ -v^{2}_{\vec{k}} $ subject to the
  constraint $  u^{2}_{\vec{k}} -v^{2}_{\vec{k}} =1 $. (2) There are very slight differences at the two peak positions,
  the latter are at  $ \pm \lbrack E^{2}(\vec{k})-\gamma _{\vec{k}}^{2}/4]^{1/2} $, while,  the former are at  $ \pm E(\vec{k} ) $.
  The analogies and differences between the $ i G^{K}_{n} $  shown in the Fig. 6 and the ARPS shown in the Fig.3 in \cite{power} can be similarly discussed.

\subsection{ Connections among the two mode squeezing spectrum, the exciton abnormal spectral weight and distribution function }

 Similarly, from Eqn.\ref{aw}, we can get the abnormal spectral weight $  \rho_{a}( \vec{k}, \omega )   =  -2 Im G^{R}_{a}( \vec{k}, \omega ) $:
\begin{eqnarray}
 \rho_{a}( \vec{k}, \omega )   &  = & - u_{\vec{k}} v_{\vec{k}}  [ \frac{ \gamma_{\vec{k}} }{ (\omega-E(\vec{k} ))^{2}+ (\gamma_{\vec{k}}/2)^{2} }
            \nonumber  \\
     & - & \frac{ \gamma_{\vec{k}} }{ (\omega + E(\vec{k} ))^{2}+ (\gamma_{\vec{k}}/2)^{2} } ]
\label{lora}
\end{eqnarray}
    where  $ u_{\vec{k}} v_{\vec{k}}= \frac{ \bar{n} V_{d}(\vec{k}) }{ 2 E(\vec{k}) } $.
    It is an odd function of $ \omega $ and satisfies the sum rule:
\begin{equation}
     \int \frac{ d \omega }{ 2 \pi}   \rho_{a}( \vec{k}, \omega ) = 0
\end{equation}
     as expected from its definition.

    It is shown in Fig.7 for several different quasi-particle energies $ E(\vec{k}) $.
    It also consists of two Lorentizan centered at $ \pm E(\vec{k} ) $ with width $ \gamma_{\vec{k}}/2 $ with the
    spectral weights $ \mp u_{\vec{k}} v_{\vec{k}} $ respectively.
    Only when $ E(\vec{k} ) \gg \gamma_{\vec{k}}/2 $, namely, when $ k \gg k^{*} $ in the Fig.3, the two Lorentizans are well separated.
    The $ \rho_{a}( \vec{k}, \omega ) $ shown in Fig.7 is an odd function of $ \omega $.
    From Eqn.\ref{kela}, one can see that the anomalous distribution function $ i G^{K}_{a} $ is a complex even function, its real and imaginary parts
    are shown in Fig.8 and Fig.9 respectively. Because the ratio of the real part over the imaginary part is independent of $ \omega $, so they have
    similar shape.

\begin{figure}
\includegraphics[width=8cm]{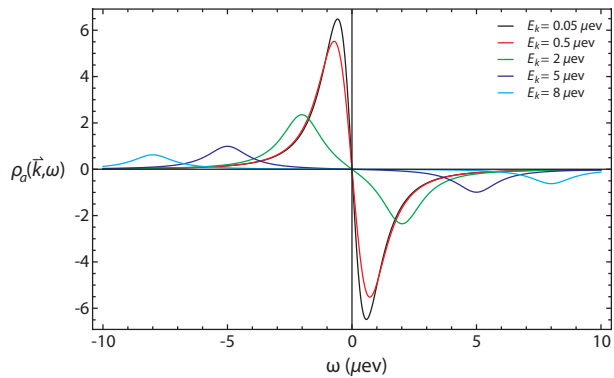}
\caption{ The anomalous  spectral weight Eqn.\ref{lora}
at different quasi-particle energies $ E( \vec{k} ) = 0.05, 0.5, 2, 5, 8 \mu e V $. }
\label{fig7}
\end{figure}

  What the second class of PL experiment measured is two mode squeezing spectrum shown in Fig.1b-3a in \cite{squ}. The excitation spectrum
  and distribution function of the exciton superfluid system to be probed by transport measurements are given by  Eqn.\ref{lora} and \ref{kela}.
  Such a relation is given by the input-output relation Eqn.\ref{inoutt},  Eqn.\ref{gla} and Eqn.\ref{lora}, Eqn.\ref{kela}.
  One can also see qualitatively the same physics with slightly quantitatively different numbers.
  When $ k < k^{*} $, the Fig.7 shows that the quasi-particle is not well
  defined. Correspondingly, the two mode squeezing spectrum in the Fig.1b in \cite{squ} has just one peak centered around $ \omega_{k}= \mu $ with the
  width $ \delta_1(\vec{k}) $ which depends on the dipole-dipole interaction $  n V_d(\vec{k}) $. The corresponding squeezing angle is always positive
  in all the frequency range as shown in Fig.4a in \cite{squ}.  From the Fig.8 and 9, one can see that the distribution function
  has just one peak  pinned at $ \omega_{k}= \mu $ with the  width $ \gamma_{k} $.
  When $ k \gg k^{*} $, the Fig.7 shows that  the two Lorentizans centered at $ \pm E(\vec{k} ) $ with the width $ \gamma_{\vec{k}}/2 $ are well separated
  with the spectral weights  $ \mp u_{\vec{k}} v_{\vec{k}} $ respectively. The quasi-particles at  $ \pm E(\vec{k} ) $ are well
  defined. Correspondingly, the  two mode squeezing spectrum in the Fig.3 in \cite{squ} has two well defined quasi-particles peaks at
  $ \omega_{k}=  \mu \pm \lbrack E^{2}(\vec{k})-\gamma _{\vec{k}}^{2}/4]^{1/2}$
  with the width $ \delta_2(\vec{k}) $ which depends on the dipole-dipole interaction.
  The corresponding squeezing angle becomes negative at small $ |  \omega_{k} -  \mu | $ and increase to be positive at large $ |  \omega_{k} -  \mu | $,
  vanishes at the resonant condition $ |  \omega_{k} -  \mu | =  \lbrack E^{2}(\vec{k})-\gamma _{\vec{k}}^{2}/4]^{1/2} $ as shown in Fig.4a in \cite{squ}.
  From the Fig.8 and 9, one can see that the distribution function split into two symmetric peaks at
  $ \omega_{k}=  \mu \pm \lbrack E^{2}(\vec{k})-\gamma _{\vec{k}}^{2}/4]^{1/2}$  with the width $ \gamma_{k}/2 $.
  It is instructive to observe some differences between
  the Fig.7 and the Fig.1b-3a in \cite{squ}: (1) The widths in the former  also depends on the the dipole-dipole interaction $  n V_d(\vec{k}) $. While
  the width in the former  is just $ \gamma_{k} $. (2) There are very slight differences at the two peak positions,
  the latter are at  $ \pm \lbrack E^{2}(\vec{k})-\gamma _{\vec{k}}^{2}/4]^{1/2} $ where the squeezing angle also vanishes,
  while, the former are at  $ \pm E(\vec{k} ) $.  The analogies and differences between the $ i G^{K}_{a} $  shown in the Fig.8 and 9
  and the Fig.1b-3a in \cite{squ} can be similarly discussed.

  Recently, the elementary excitation spectrum of
  exciton-polariton inside a micro-cavity was measured \cite{expolmode} and
  was found to be very similar to that in a $^{4}He $ superfluid except in a small regime near $ k=0 $. We believe this
  observation on the anomaly near $ k=0 $ is precisely due to the excitation spectrum in a
  non-equilibrium stationary superfluid shown in Fig.5 and Fig.7.

\begin{figure}
\includegraphics[width=8cm]{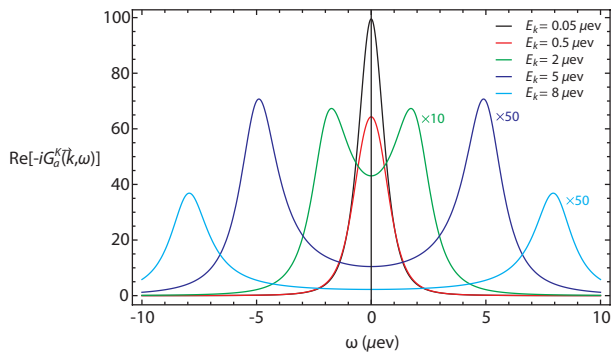}
\hspace{0.3cm}
\caption{ The real part of the anomalous Keldysh component $ i G^{K}_{a} $ at different quasi-particle energies $ E( \vec{k} ) = 0.05, 0.5, 2, 5, 8 \mu e V $.}
\label{fig8}
\end{figure}

\begin{figure}
\includegraphics[width=8cm]{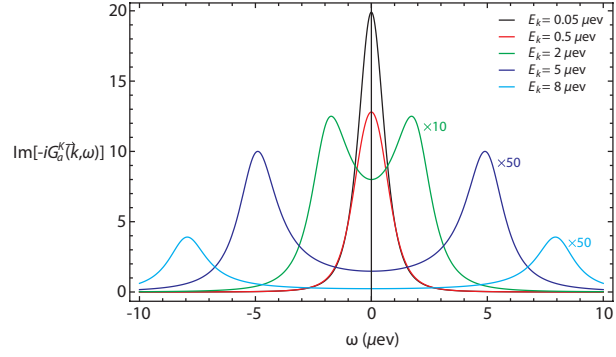}
\hspace{0.3cm}
\caption{ The imaginary part of the anomalous Keldysh component $ i G^{K}_{a} $ at different quasi-particle energies $ E( \vec{k} ) = 0.05, 0.5, 2, 5, 8 \mu e V $. }
\label{fig9}
\end{figure}

\section{ Two exciton correlation functions  }

In this section, we will explore the relations between the two photon correlations measurements and the two exciton correlation functions.
Similar to two photon correlation functions studied in \cite{squ}, one can also define the two exciton
correlations functions:
\begin{eqnarray}
g_{2b}^{(\vec{k})}(\tau )& = & \frac{\left\langle b_{\vec{k}}^{\dagger }(t)b_{
\vec{k}}^{\dagger }(t+\tau )b_{\vec{k}}(t+\tau )b_{\vec{k}}(t)\right\rangle_{in} }{\left\vert G_{1b}(0)\right\vert ^{2}}
    \nonumber  \\
g_{2b}^{(\pm \vec{k} )}(\tau ) & = & \frac{\left\langle
b_{\vec{k}}^{\dagger }(t+\tau
)b_{\vec{k}}(t+\tau )b_{-\vec{k}}^{\dagger }(t)b_{-\vec{k}%
}(t)\right\rangle_{in} }{\left\vert G_{1b}(0)\right\vert ^{2}}
\label{g2bboth}
\end{eqnarray}
where the $ G_{1b}(\tau)= \langle b_{\vec{k}}^{\dagger }(t+\tau )
b_{\vec{k} }(t) \rangle_{in} $ is the single exciton correlation
function in Eqn.\ref{g1b} and shown in the Fig.4.
In order to directly contrast with the  two photon correlation functions calculated in \cite{squ},
we first consider the orderings of the exciton opertaors in Eqn.\ref{g2bboth}. The other orders will be discussed near to the end of this section.

Due to the input-output relation Eqn.\ref{inout}, one can see
immediately that $ g_{2b}^{(\vec{k})}(\tau )= g_{2}^{(\vec{k})}(\tau
) $ which is given in Eqn.12 in \cite{squ}. However, due to the highly
non-trivial input-output relation Eqn.\ref{inoutt}, the $ g_{2b}^{(\pm
\vec{k} )}(\tau ) $ could be even qualitatively differ from $
g_{2}^{(\pm \vec{k} )}(\tau ) $ given in Eqn.12 in \cite{squ}. This is indeed the case as shown below.

By using Eqn.\ref{b} and the commutation relations in the frequency
space $ [ a_{\vec{k} }^{in}( \omega ), a_{\vec{k} }^{in \dagger }(
\omega^{\prime} ) ]= \delta(\omega-\omega^{\prime} ) $, one can find
\begin{equation}
G_{2b}^{( \pm \vec{k})}(\tau )  =  G^{2}_{1b}(\vec{k}, \tau=0 )+
F_{1b}( \tau) \times F^{*}_{2b}( \tau ) \label{g2pmbige}
\end{equation}
   where the $ F_{1b}(\tau) $ and $ F_{2b}(\tau) $ are given in
   Eqn.\ref{f1f2b}. In fact, one can
physically understand Eqn.\ref{g2pmbige} by noting that the first
term is due to the equal time single exciton correlation function $
 G_{1b}(\tau=0)=\langle b_{ \pm \vec{k}}^{\dagger }(t ) b_{ \pm \vec{k} }(t) \rangle
$ in the Eqn.\ref{g1b}, while the second term is due to the exciton
anomalous Green function Eqn.\ref{gga}  which encodes the exciton phase correlations.
 Because $ F_{1b}(\tau) \neq F_{2b}(\tau) $, so the $ G_{2b}^{(\vec{k})}(\tau ) $ must be   complex.
 Then the normalized two exciton correlation $ g_{2b}^{(\pm \vec{k} )}(\tau )=1+  f_{1b}(\tau
) \times f^{*}_{2b}( \tau ) $ where the $ f_{1b}(\tau ) $ and $
f_{2b}(\tau ) $ are given by Eqn.\ref{f1f2b}, so
\begin{eqnarray}
g_{2b}^{(\pm \vec{k} )}(\tau ) &=&  g_{2}^{(\pm \vec{k} )}(\tau ) -i
2 (\frac{E^{2}(\vec{k})+\frac{\gamma
_{\vec{k}}^{2}}{4}}{\bar{n}V_{d}(\vec{k})})^{2}e^{-\gamma_{\vec{k}}\tau
} \frac{ \sin(E(\vec{k})\tau )}{E(\vec{k}) }  \nonumber  \\
& \times &  [u_{\vec{k}}^{2}\frac{e^{-iE(\vec{k})\tau
}}{E(\vec{k})-i\frac{\gamma
_{\vec{k}}}{2}}+v_{\vec{k}}^{2}\frac{e^{iE(\vec{k})\tau
}}{E(\vec{k})+i\frac{ \gamma _{\vec{k}}}{2}}] \label{gpmb}
\end{eqnarray}

  The HanburyBrown-Twiss type of experiments are designed to measure the two photon correlation function shown in Fig.3b in \cite{squ}.
  The two exciton correlation  function  of the exciton superfluid system is given by  Eqn.\ref{gpmb}.
  Although the normalized  two photon correlation function $ g_{2}^{(\pm
\vec{k} )}(\tau )$ ( shown in Eqn.12 and Fig.3b in \cite{squ} ) is always positive, due to the
complex second term in Eqn.\ref{gpmb}, the $ g_{2b}^{(\pm \vec{k})}(\tau )$ remains complex. We can separately evaluate
both the real part and the imaginary part of the difference of the two:
\begin{eqnarray}
Re g_{2b}^{(\pm \vec{k} )}(\tau ) & - & g_{2}^{(\pm \vec{k} )}(\tau ) = -
2 \frac{ E^{2}(\vec{k})+ (\frac{\gamma_{\vec{k}}}{2})^{2} } { ( \bar{n}V_{d}(\vec{k}) )^{2} } e^{-\gamma_{\vec{k}}\tau }  \nonumber  \\
& \times & [ \sin^{2}E(\vec{k})\tau -  \frac{\gamma_{\vec{k}}}{4 E(\vec{k})} \sin 2 E(\vec{k})\tau ]   \nonumber  \\
Im g_{2b}^{(\pm \vec{k} )}(\tau ) & = & -
 \frac{ E^{2}(\vec{k})+ (\frac{\gamma_{\vec{k}}}{2})^{2} } { ( \bar{n}V_{d}(\vec{k}) )^{2} } e^{-\gamma_{\vec{k}}\tau }
( u^{2}_{\vec{k}}+ v^{2}_{\vec{k}})   \nonumber  \\
& \times & [ \sin 2E(\vec{k})\tau  +  \frac{\gamma_{\vec{k}}}{E(\vec{k})} \sin^{2} E(\vec{k})\tau ]
\label{twoexreim}
\end{eqnarray}
   which is drawn in the Fig.10.

\begin{figure}
\includegraphics[width=8cm]{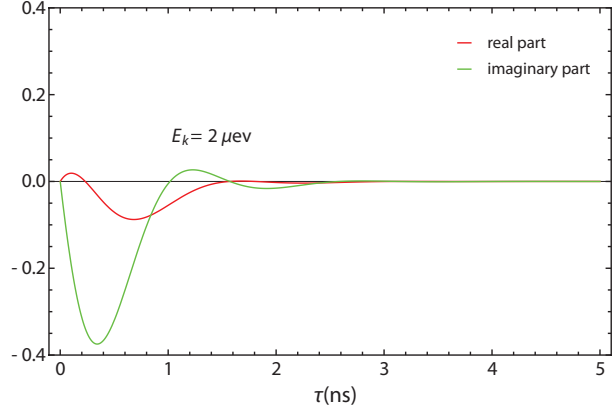}
\caption{ Real and Imaginary part of the two exciton correlation functions Eqn.\ref{twoexreim} for $ E( \vec{k} ) =  2 \mu e V $.}
\label{fig10}
\end{figure}

   When comparing the two exciton correlation function with the single exciton correlation function in Eqn.\ref{g1b} and
   shown in the Fig.4, one can see that the envelop function in the former ( latter ) decays with the $ \gamma_{\vec{k}} $ (  $ \gamma_{\vec{k}}/2 $ ),
   the oscillation frequency in the former ( latter ) is  $ 2 E(\vec{k})$ (  $  E(\vec{k}) $ ).

   One can see why $ g_{2b}^{(\vec{k})}(\tau ) $ in Eqn.\ref{g2bboth} must be positive definite.
   Define $ A^{\dagger}= b_{\vec{k}}^{\dagger }(t)b^{\dagger }_{\vec{k}}(t+\tau ),
   A=b_{\vec{k}}(t+\tau)b_{\vec{k} }(t) $, so the numerator can be written as $ H= A^{\dagger} A $ which, of course,
   is a positive definite Hermitian operator.
   Similar arguments can be used used to show the two photon correlation functions defined in \cite{squ} must be positive definite.
   In fact, one can calculate various two exciton correlation functions by  shifting the orders of the exciton operators or
   put Time-ordered inside the Eqn.\ref{g2bboth}. For example, one can consider
   $ \left\langle b_{\vec{k}}^{\dagger }(t+\tau )b^{\dagger}_{-\vec{k}}(t) b_{-\vec{k}}(t)b_{\vec{k} }(t+\tau)  \right\rangle_{in} $,
   if one define $ A^{\dagger}= b_{\vec{k}}^{\dagger }(t+\tau )b^{\dagger}_{-\vec{k}}(t), A=b_{-\vec{k}}(t)b_{\vec{k} }(t+\tau) $, the argument
   can be written as $  H= A^{\dagger} A $, so it must be  positive definite. Similar procedures as above lead to its value
   $ 1 + | f_{1b}(\tau ) |^{2} =  g_{2}^{(\pm \vec{k} )}(\tau ) $. One can show that the other two orderings lead to
   $  1 + | f_{2b}(\tau ) |^{2} $ which is positive definite  and  $ 1+  f^{*}_{1b}(\tau) \times f_{2b}( \tau ) =  g_{2b}^{*(\pm \vec{k} )}(\tau ) $
   which is complex conjugate to Eqn.\ref{gpmb} and shown in the Fig.10 by adding a minus sign to the imaginary part.

   Various exciton correlation functions discussed in this section are needed to evaluate the transport properties of excitons such the Coulomb drag and
   the counterflow resistances.

\section{ Exciton correlation functions in an  equilibrium  dissipative exciton superfluid }

  In this section, we will study the Exciton correlation functions in an dissipative equilibrium  exciton superfluid in Fig.2
  and then compare with those in the stationary non-equilibrium exciton superfluid in Fig.1 studied in all the previous sections.

  As shown in the Sec.II, in the rotating frame with the frequency $ \mu $, the quantum open system in Fig.1 has the Hamiltonian Eqn.\ref{firstr} and
  the effective quadratic Hamiltonian Eqn.\ref{quad}. In the lab frame, the equilibrium system in Fig.2 have the same Hamiltonian.
  Staring from Eqn.\ref{firstr} and effective quadratic Hamiltonian Eqn.\ref{quad}, we will discuss $ \vec{k}=0 $ and  $ \vec{k} \neq 0 $ respectively.

   At $ \vec{k}=0 $, the Hamiltonian for the photons are:
\begin{eqnarray}
   H_{0} & = &  \sum_{k_z} ( v_g | k_z | - \mu )  a^{\dagger}( k_z ) a( k_z )
                         \nonumber  \\
   & + & i \sum_{k_z} \sqrt{N} ( g(k_z)  a -  g^{*}(k_z) a^{\dagger} )
\end{eqnarray}
    Obviously, it can be rewritten as
\begin{eqnarray}
   H_{0}  =  \sum_{k_z} ( v_g | k_z | - \mu)  ( a^{\dagger}( k_z )- \langle a( k_z ) \rangle^{*} )( a( k_z )-\langle a( k_z ) \rangle ) + const.
                                 \nonumber  \\
   \langle a( k_z ) \rangle  =   \frac{ i g(k_z) \sqrt{N} }{  v_g | k_z | - \mu } ~~~~~~~~~~~~~~~~~~~~~~~~
\end{eqnarray}
    where $   \langle a( k_z ) \rangle $ is the expectation value of the photon annihilation operator in the ground state.
    Because the coupling constant $  g(k_z) \sim L^{-1/2}_{z} \rightarrow 0 $ in the $ L_{z} \rightarrow \infty $ limit,
    so a proper regulization is needed to extract the physics embedded in this expectation value. This was done in \cite{power}.
    Following the regulization procedure in \cite{power}, we can find the average photon number in the ground state
\begin{equation}
     n_{\omega _{k_{z}}}= N \gamma_{0} \delta( v_g | k_z | - \mu)
\label{photoncon}
\end{equation}
 where the $ \gamma_0 =|g|^{2}D $ is the exciton decay rate at $ \vec{k}=0 $ where
    the $ D= L_{z}/v_{g} $ is the  photon density of states at $ \vec{k}=0 $.
    One can see that the $ \gamma_0 $  is independent of $ L_{z} $ and is an experimentally measurable quantity \cite{power}.
    The photon condensation  Eqn.\ref{photoncon} is the same as the emitted photon number at $ \vec{k} = 0 $ Eqn.4 in Ref.\ref{power}.

    Now we will calculate the exciton correlations at $ \vec{k} \neq 0 $ in the ground state of the interacting system.
    Obviously, the exact ground state of the interacting systems $ | G \rangle $ in Fig.2 is not the same as the initial input state in Eqn.\ref{in}:
\begin{equation}
| G \rangle \neq |in\rangle = |BEC\rangle |0\rangle_{ph}
\label{exactg}
\end{equation}%
    Even we do not know the explicit form of $ | G \rangle $, we are still  able to calculate all the ground state correlation
    functions by the imaginary time path-integral method which is particularly suitable to calculate correlation functions in
    an equilibrium system \cite{scaling}.

    In the imaginary time path integral representation of the Hamiltonian Eqn.\ref{first}
\begin{eqnarray}
    {\cal L} &= & \sum_{\vec{k}} \int^{\beta}_{0} d \tau [ b^{\dagger}_{\vec{k}}(\tau) \partial_{\tau} b_{\vec{k}}(\tau)
    +   a^{\dagger}_{k}(\tau) \partial_{\tau} a_{k}(\tau)  ] \nonumber  \\
    & + & \int^{\beta}_{0} d \tau (   H_{sf} + H_{ph}+H_{int}  )
\label{path}
\end{eqnarray}

    Because the total action Eqn.\ref{path} is quadratic, integrating out the photons out exactly leads to the effective  action for the excitons only:
\begin{eqnarray}
    {\cal L}_{ex} & = & \frac{1}{\beta} \sum_{ i \omega_n} \sum_{\vec{k}} [ b^{\dagger}_{\vec{k}}( i \omega_n )
    ( -i  \omega_n + \Sigma( \vec{k}, i \omega_n ) )  b_{\vec{k}}( i \omega_n )  \nonumber  \\
    & + & (\epsilon _{\vec{k}}+V_{d}(\vec{k})\bar{n}) b_{
    \vec{k}}^{\dagger }( i \omega_n ) b_{\vec{k}}( i \omega_n )     \nonumber  \\
    & + & (\frac{V_{d}(\vec{k})\bar{n}}{2}
    b_{\vec{k}}^{\dagger }( i \omega_n )  b_{-\vec{k}}^{\dagger }( -i \omega_n )+ h.c.)]
\label{actionex}
\end{eqnarray}
     where the $ \Sigma( \vec{k}, i \omega_n ) $ is the self energy due to the integration of photons:
\begin{equation}
     \Sigma( \vec{k}, i \omega_n )= \sum_{k_z}  \frac{ |g(k)|^{2} }{ i \omega_n -( \omega_k -\mu) }
\label{self}
\end{equation}
     where the sum is over continuous modes of photons at a given in-plane momentum $ \vec{k} $.

     One can easily find the {\sl imaginary time ordered $ {\cal T } $ } normal and anomalous exciton correlation functions \cite{justify}:
\begin{eqnarray}
{\cal G}^{{\cal T}}_{n}(\vec{k}, i\omega_n ) & =  & \frac{ i\omega_n -\Sigma( \vec{k}, i \omega_n )  +\epsilon _{\vec{k}}+\bar{n}V_{d}( \vec{k})}
{ ( i\omega_n -\Sigma( \vec{k}, i \omega_n )) ^{2}-E^{2}(\vec{k})}   \nonumber   \\
{\cal G}^{{\cal T}}_{a}(\vec{k},\omega ) & =  &  \frac{ - \bar{n}V_{d}(\vec{k})}{ ( i\omega_n -\Sigma( \vec{k}, i \omega_n )) ^{2}-E^{2}(\vec{k})}
\label{it}
\end{eqnarray}
     where the average is with respect to the exact ground state of the system. Obviously, the initial input state Eqn.\ref{in} is not even an eigen-state
     of the system, let alone the ground state. This is the main difference between the equilibrium correlation functions calculated here
     and those calculated in the previous sections. It is the goal of this section to contrast some analogies and especially, the crucial
     differences between this two sets of correlation functions.

     After doing the analytic continuation $ i \omega_n \rightarrow \omega + i \eta $ in Eqn.\ref{self} and replacing the sum by
     an integral over the density of state $  \sum_{k_z} \rightarrow  \int d \omega_{k}  D_{\vec{k}}( \omega_{k} ) $, one can see:
\begin{equation}
     \Sigma( \vec{k}, \omega + i \eta )= P \int d \omega_{k} \frac{ D_{\vec{k}}( \omega_{k} ) |g(k)|^{2} }{ \omega -( \omega_k -\mu) } + i \gamma_k/2
\label{selfri}
\end{equation}
     where the $ P $ stands for the principle part of the integral and the exciton decay rate $ \gamma_k  $ is:
\begin{eqnarray}
     \gamma_k  & = &  2 \pi \int d \omega_{k}  D_{\vec{k}}( \omega_{k} ) |g(k)|^{2} \delta( \omega -( \omega_k -\mu) )   \nonumber  \\
     & \sim & 2 \pi D_{\vec{k}}(\mu )\left\vert g_{\vec{k}}(\omega _{k}=\mu )\right\vert ^{2}
\label{rate}
\end{eqnarray}
     where we made the Markov approximation used in \cite{power,squ}. The real part causes a small energy shift
     which, in fact, vanishes identically within the  Markov approximation in \cite{power,squ}. Substituting Eqns.\ref{selfri} and \ref{rate} into Eqn.\ref{it} leads to:
\begin{eqnarray}
  {\cal G}^{R}_{n}(\vec{k},\omega ) & = & G_{n}(\vec{k},\omega + i\frac{\gamma _{k}}{2})={\cal G}^{A*}_{n}(\vec{k},\omega )
  \nonumber  \\
  {\cal G}^{R}_{a}(\vec{k},\omega ) & = & G_{a}(\vec{k},\omega + i\frac{\gamma _{k}}{2})={\cal G}^{A*}_{a}(\vec{k},\omega )
\label{ran2e}
\end{eqnarray}
     which are identical to Eqn.\ref{ran2} and Eqn.\ref{raa2}.
     At first sight, this result may look surprising. However, one can understand it better
     by comparing the equilibrium dissipative exciton superfluid in Fig.2 discussed in this section
     with the non-equilibrium exciton superfluid system in Fig.1 discussed in the previous sections.
     Both systems are described by the same Hamiltonian Eqn.\ref{first} and \ref{quad}, so should have the same excitation spectrum.
     This leads to Eqn.\ref{ran2e} from a very general physical picture. However, what distinguishes the two systems is the initial conditions.
     Only the distribution function, namely, the Keldysh component $ G^{K} $ depends on the initial conditions. While the
     retarded and advanced Green functions $ G^{R} $ and $ G^{A} $ depends only on the Hamiltonian.
     The distribution function for a bosonic equilibrium system is simply given by the
     Bose distribution function, while that for an non-equilibrium system is determined by the initial conditions, namely the input state Eqn.\ref{in}.

   The greater and lesser normal Green functions at $ T=0 $ can also be found from the Fluctuation and Dissipation Theorem (FDT):
\begin{eqnarray}
   i {\cal G}^{>}_{n}(\vec{k},\omega ) & = & \theta ( \omega) \rho_{n}(\vec{k},\omega )
  \nonumber  \\
   i {\cal G}^{<}_{n}(\vec{k},\omega ) & = & - \theta ( -\omega) \rho_{n}(\vec{k},\omega )
\label{glgne}
\end{eqnarray}

     Obviously, they differ from the corresponding Green functions in the non-equilibrium case Eqn.\ref{ggn},\ref{gln}.
     Of course, we still have the identity:
\begin{equation}
     {\cal G}^{R}_{n}- {\cal G}^{A}_{n}={\cal G}^{>}_{n}-{\cal G}^{<}_{n}
\end{equation}

     The real time-ordered $ T $ normal Green function can be found:
\begin{eqnarray}
    {\cal G}^{T}_{n}(\vec{k},\omega )  & = & {\cal G}^{R}_{n}(\vec{k},\omega ) \theta(\omega)+ {\cal G}^{A}_{n}(\vec{k},\omega ) \theta(-\omega)
                                \nonumber \\
    &  = &  G_{n}(\vec{k},\omega + i\frac{\gamma _{k}}{2} sgn \omega )
\label{togne}
\end{eqnarray}
    Obviously, it differs from the corresponding Green function in the non-equilibrium case Eqn.\ref{gtn}.

    From Eqn.\ref{glgne}, one can get the Keldysh component Green function:
\begin{eqnarray}
    i {\cal G}^{K}_{n}(\vec{k},\omega ) = i  ( {\cal G}^{>}_{n}(\vec{k},\omega ) + {\cal G}^{<}_{n}(\vec{k},\omega ) )  \nonumber  \\
    = sgn (\omega) \rho_{n}(\vec{k},\omega ) = - 2 Im {\cal G}^{T}_{n}(\vec{k},\omega )
\label{keldeq}
\end{eqnarray}
     which is nothing but the FDT Eqn.\ref{keln} in any equilibrium system. Obviously, it is quite different than $ i G^{K}_{n} $ in Eqn.\ref{keln}
     for the non-equilibrium system.

    Very similar expressions for  the greater, the lesser, time ordered anomalous Green functions and the Keldysh component
    can be written down by changing the subscript $ n $
    from Eqn. \ref{glgne} to Eqn.\ref{keldeq} to the  subscript to $ a $.
    Two exciton correlation functions can be computed using the Wick theorems. Their analogies and differences with those of the excitons in
    non-equilibrium systems calculated in Sec. VIII can be addressed similarly.

     Because the retarded and advanced Green functions are the same as those in the non-equilibrium case,
     so the normal and anomalous spectral weights are the same as those in the non-equilibrium case listed
     in Eqn.\ref{lorn}, \ref{lora} and drawn in the Figs.5 and 7. However, the distribution functions are quite different.
     For the equilibrium case in the Fig.2, the distribution functions are just multiply Figs.5 and 7 by $ sgn(\omega) $.
     While for the non-equilibrium case in the Fig.1, the distribution functions are shown in the Fig.6,8 and 9.

\section{Conclusions}

  In this paper, by using the Heisenberg-Langevin equations in the context of the  input-output formalism,
  we calculated various one and two exciton correlation functions
  and also explored their relations to various photon correlation functions.
  We evaluated both the normal and abnormal spectral weights which lead to the excitation spectra of the non-equilibrium exciton superfluids.
  We also studied both the normal and abnormal Kelydsh component Green functions which lead to the
  distribution functions of the non-equilibrium exciton superfluids.
  These relations brought out the important connections between the properties of exciton superfluid systems to be probed by transport easurements and
  the various quantum optical photoluminescence measurements on emitted photons such as the angle resolved photon power spectrum,
  phase sensitive homodyne experiments and HanburyBrown-Twiss type of experiments.

  By the imaginary time formalism, we calculated the exciton correlations functions of an equilibrium exciton superfluid
  subject to a photon dissipation bath. It was demonstrated explicitly that $ G^{R}_{n/a}, G^{A}_{n/a} $ and the spectral weights $ \rho_{n/a} $ are the same
  for the two systems, but the $ G^{>}_{n/a}, G^{<}_{n/a} $, $ G^{T}_{n/a} $ and
  the distribution functions, namely, the Keldysh component Green functions $ G^{K}_{n/a} $ are different.
  The analogies and differences between the two kinds of systems can be summarized in the Fig.1 and Fig.2.
  In the equilibrium system in Fig.2, one measures $ {\cal G}^{<} $  experimentally which is directly related to the
  spectral weight $ \rho $ by the FDT Eqn.\ref{glgne}.
  ( see also \cite{scaling} ).
  The conventional theoretical tool is the imaginary time path integral method which can be analytically continued to  the real time as shown in the Sec.IX.
  While in an non-equilibrium system in Fig.1, in the present context of non-equilibrium exciton superfluids, one also measures
  $ G^{<}_{ph} $ for emitting photons
  which is related to the corresponding  $ G^{<} $ for the excitons by the input-output relation Eqn.\ref{inout} derived in Section VI.
  Unfortunately, due to the lack of the FDT, the $  G^{<} $ for the excitons is not simply related to the spectral weight $ \rho $,
  its relation is non-trivial and need to be studied in details. For the given initial conditions Eqn.\ref{in},
  this relation has been worked out by the Heisenberg equation of motions and was discussed explicitly in Sec.IV-VII.
  This important feature should hold in any non-equilibrium quantum open system.

  The input-output formalism can only be used to study the dynamic scattering process in Fig.1 starting from the input initial
  state Eqn.\ref{in}.
  While the imaginary time path integral method in the section IX can only be used to study the equilibrium case in Fig.2 based on the exact ground state Eqn.\ref{exactg}. It was known that the correlation functions in both equilibrium and non-equilibrium quantum open systems
  could also be calculated by the Keldysh Green function method.
  The Kelydsh formalism in either Canonical Quantization language or path-integral language can be used to study both
  the dynamic scattering process in Fig.1 starting from the input initial state Eqn.\ref{in} and the
  the equilibrium case in Fig.2 based on the exact ground state Eqn.\ref{exactg}.
  In a non-equilibrium case,  because the $ G^{>}_{n/a} $ and $ G^{<}_{n/a} $ are independent of each other,
  one need both the forward and the backward paths in the real-time Keldysh contour.
  While in an equilibrium case, they are related by Fluctuation and dissipation theorem (FDT), so
  the  forward and backward path can be reduced to just the forward path.
  In the Kelydsh formalism, one can find $ G^{R}_{n/a}, G^{A}_{n/a} $, therefore the spectral weight
  $ \rho_{n/a}=  i ( G^{R}_{n/a}- G^{A}_{n/a}) = i ( G^{>}_{n/a} - G^{<}_{n/a} ) $ which leads to excitation spectrum of
 the interacting system. Most importantly, one can also determine the
 Keldysh component  $ G^{K}_{n/a} =  G^{>}_{n/a} + G^{<}_{n/a} $ which leads to the distribution function.
 It is the   distribution function which lead to the difference between
 the equilibrium and the non- equilibrium system. Then one can determine  $ G^{>}_{n/a} $ and $ G^{<}_{n/a} $ respectively
 and compare with those achieved from
 the input-output formalism for non-equilibrium system and  the imaginary time path integral method for the equilibrium system.
 Two-exciton correlation functions can also be evaluated by the Keldysh formalism by the counter-ordered Wick theorem.
 As shown in the Sect.IV-VIII, the procedures using the input-output formalism  is just opposite to those using the Kelydsh formalism:
 one  determine  $ G^{>}_{n/a} $ and $ G^{<}_{n/a} $ first,
 then the spectral weight  $ \rho_{n/a}=  i ( G^{>}_{n/a} - G^{<}_{n/a} ) $, then determine $ G^{R}_{n/a}, G^{A}_{n/a}, G^{T}_{n/a} $ and  $ G^{K}_{n/a} $.
 The connections between the Keldysh Green function formalism and the
 input-output formalism will be explored in details in a future publication \cite{sun}.

 The various one and two exciton correlation functions can be used to calculate
 the transport properties of the excitons such as the Coulomb drag and counterflow experiments.
 They are also needed to calculate the superfluid density and the critical velocity of the dissipative superfluids.
 Indeed, as pointed out in the introduction, the transport properties can provide more direct evidences on superfluids than the spectroscopic properties.
 The input-output relations can be used to establish the connections between
 the transport measurements on excitons and the PL measurements. In a future publication, using all the one and two exciton correlation functions
 derived in this paper, we will compute the Coulomb drags, the counterflow resistances, superfluid densities and critical velocities on both equilibrium dissipative superfluids in the Fig.2 and the non-equilibrium open system in the Fig.1.

   As pointed out in the introduction, in this paper, we ignored the spins of the electrons and holes, therefore also the polarization
 of emitted photons. In fact,
 the electrons in the conduction band carry spin $ s= \pm 1/2 $. Due to the
 spin-orbit coupling, the holes in the valence band carry total spin $ J =3/2 $. The quantum well
 confinement potential in the EHBL splits the energy of the light hole with $ m_{l}= \pm 1/2 $ above
 that of the heavy hole $ m_{h}=\pm 3/2 $ \cite{spinthe}. So we will neglect the
 higher energy light-hole, only consider the lower energy heavy-
 hole with $ m_{h}=\pm 3/2 $.  One  electron with $ s= \pm 1/2 $ and one heavy-hole
 with $ m_{h}=\pm 3/2 $ can bind into four kinds of
 heavy-hole excitons $ (\pm 1/2, \pm 3/2) $.
 When coupled to photons, the selection rule of the photon angular
 momentum is $ J= s+ m_{h} $, so the 4 kinds of heavy-hole excitons can be
 grouped into the bight excitons  $ J= \pm 1 $ which couple to the one photon
 process with the polarization $ \sigma=\pm $ and the dark excitons  $ J=  \pm 2 $ which do not couple to the one
 photon process. However, the spins of electrons and holes inside an
   exciton will relax ( or de-cohere ) due to the spin-orbit couplings in the
   conduction band and valence band respectively.
   These individual electron and hole spin flips lead to the conversion from the dark excitons to the
   bright excitons. The simultaneous flipping of both electrons and holes is due
   to the Coulomb exchange interaction, it leads to a spin flip between $ J=1 $ and $ J=-1 $ inside the bright
   excitons. In order to make quantitative connection between theoretical predictions with
   the experiments, it would be important to study how the results achieved in \cite{power,squ} and in this paper
   for spinless excitons will be modified after taking into account of both bright and dark excitons, also the effects of a trap, disorder and
   finite temperatures.

\vspace{1cm}

{\bf Acknowledgements }

 J.Ye thank T. Shi and Longhua Jiang for the collaborations on the previous works \cite{power,squ} which are complementary to the present work.
 J. Ye's research is supported by NSF-DMR-1161497, NSFC-11074173, NSFC-11174210, Beijing Municipal Commission of Education under grant No.PHR201107121,
 at KITP is supported in part by the NSF under grant No. PHY11-25915. W.M. Liu's research was supported by NSFC-10874235.


\begin{thebibliography}{99}
\bibitem{factori} In order to be consistent with the conventional notations in the Green functions in condensed matter system \cite{fetter},
 compared to the notations used in \cite{power,squ}, we made a change $ G_{n} \rightarrow i  G_{n},  G_{a} \rightarrow -i  G_{a} $.
\bibitem{tilde} In the following, we negelct the $ \tilde{} $ on top of the exciton opertaor.
\bibitem{diff} In order to show clearly the different structures of various correlation functions, we choose different values than those used in
\cite{power,squ}.
\bibitem{draw} When drawing Fig.5-9, for given $   n V_d(\vec{k}) =5 \mu e V, \gamma_{\vec{k}}/2= 1 \mu e V $ and various $ E( \vec{k} ) $, we use
 $ u_{\vec{k}}^{2}  =\frac{ \sqrt{ E^{2}(\vec{k})+ (\bar{n}V_{d}(\vec{k}))^{2} }}{2E(\vec{k%
})} + \frac{1}{2},  v_{\vec{k}}^{2} = \frac{ \sqrt{ E^{2}(\vec{k})+ (\bar{n}V_{d}(\vec{k}))^{2} }}{2E(\vec{k%
})} - \frac{1}{2}  $ to determine $ \rho_n $ in Eqn.\ref{lorn} in Fig.5,
 $ \rho_a $ in Eqn.\ref{lora} in Fig.7 and Eqn.\ref{twoexreim} in Fig.10.
\bibitem{justify} Here we assume $ \Sigma( - \vec{k}, - i \omega_n ) = - \Sigma(  \vec{k}, i \omega_n ) $. This is indeed justified
under the Markov approximation Eqn.\ref{rate}. the crucial point here is the existence of the large optical characteristic frequency $ \mu $.
However, in the dissipative exciton superfluids in BLQH \cite{blqhexp,counterflow,psdw}, there is gapless
electron-hole excitations instaed of such a large characteristic frequency, so the
Markov approximation never holds in this kind of system. This is also the cruucial differences between the dissipations in quantum optical systems
and in condesed matter systems.
\end{thebibliography}
\end{document}